\documentclass[10pt, doublecolumn]{IEEEtran}

\usepackage{epsfig,latexsym}
\usepackage{float}
\usepackage{indentfirst}
\usepackage{amsmath}
\usepackage{bm}
\usepackage{amssymb}
\usepackage{times}
\usepackage{graphicx}
\usepackage[ruled,linesnumbered,boxed]{algorithm2e}
\usepackage[noend]{algpseudocode}
\usepackage{subfigure}
\usepackage{psfrag}
\usepackage{hyperref}
\usepackage{cite}
\usepackage{lastpage}
\usepackage{fancyhdr}
\usepackage{color}
 \usepackage{amsthm}
\usepackage{bigints}
\newcounter{problem}
\newcounter{save@equation}
\newcounter{save@problem}
\makeatletter

\begin{document}
\title{\vspace{-0.5em} \huge{Priority Random Access and Power Control for NOMA-ALOHA in Heterogeneous mMTC}}

\author{
Wenbo Fan, \IEEEmembership{Graduate Student Member, IEEE},
Pingzhi Fan, \IEEEmembership{Life Fellow, IEEE}, \\ and
Zilong Liu, \IEEEmembership{Senior Member, IEEE}

\thanks{ Wenbo Fan and Pingzhi Fan are with the Information Coding  \& Transmission Key Lab of Sichuan Province, CSNMT Int. Coop. Res. Centre (MoST), Southwest Jiaotong University, Chengdu, China (e-mail: \href{mailto:fwb@my.swjtu.edu.cn}{fwb@my.swjtu.edu.cn};
	\href{mailto:pzfan@swjtu.edu.cn}{pzfan@swjtu.edu.cn}).}
\thanks{ Zilong Liu is with the School of Computer Science and Electronic Engineering, University of Essex, Colchester CO4 3SQ, U.K. (e-mail: \href{mailto:zilong.liu@essex.ac.uk}{zilong.liu@essex.ac.uk}).}

 \vspace{-1em}}
 \maketitle
 
\vspace{-1em}
\begin{abstract}
  This paper presents a novel priority random access (PRA) non-orthogonal multiple access assisted ALOHA, called PRA-NA, to provide access priority for machine-type devices (MTDs) with different delay requirements (i.e., delay-sensitive and delay-tolerant). We first introduce a received power level model that incorporates imperfect channel state information and imperfect successive interference cancellation to study the impact of practical non-ideal channel conditions. Two PRA strategies including fixed PRA-NA (FPRA-NA) and adaptive PRA-NA (APRA-NA) are then designed to reduce the average access delay of delay-sensitive MTDs in heterogeneous massive machine-type communications. Subsequently, the throughputs of both the FPRA-NA and APRA-NA strategies are analyzed to demonstrate their effectiveness. Moreover, to improve the energy efficiency of random access, we introduce an enhanced user barring algorithm (EUBA) to carry out  power control. It is shown that our proposed EUBA can not only alleviate the user overload problem, but also reduce the average transmit power of MTDs. By extending it to the proposed PRA-NA schemes, we demonstrate via extensive simulation results that the random access performances in terms of throughput, access delay, and energy efficiency can be significantly improved over the conventional NOMA-ALOHA.
 
\end{abstract}


 \section{Introduction}
 
 \subsection{Background}
 
With the wide proliferation of machine-type devices (MTDs) in the Internet of Things (IoT), massive machine-type communications (mMTC) have been identified as a critical use case for the next-generation mobile systems. Different from human-centric communications, mMTC are expected to support a connection density of 10 million per kilometer square in the sixth-generation (6G) mobile systems, according to the third-generation partnership project (3GPP) reports \cite{8766143,1903968}. 

Typically, mMTC data services are characterized by sparse and sporadic transmission with short packets (e.g., hundreds of bits) \cite{9170262,10037184}. Often, one needs to support a large set of uncoordinated devices concurrently communicating over the shared wireless medium. A burning problem of such type of transmission is that the coordination of these randomly emitted signals may consume significant signaling overhead. To address such a problem, there has been a renewed interest for studying random access technologies in recent years \cite{9057670}. 

Cellular-based IoT systems (e.g., NarrowBand-IoT and  Long-Term Evolution (LTE)-Machine to Machine) are considered as a promising solution for mMTC scenarios due to their wide coverage, high reliability, and mobile support. However, in the contention-based random access for conventional cellular networks, due to the limited preamble resources in random access channel (RACH), serious user collision may occur during the four-step random access process when there are a large number of MTDs \cite{1611434}. To alleviate the user overload problem, several random access control schemes have been proposed including access class barring (ACB) \cite{7404058,7875393}, back-off (BO) \cite{9013247}, and hybrid ACB-BO \cite{9210822}. Although these schemes can help to improve the system throughput of the grant-based random access, it is challenging to support a massive number of MTDs due to limited preamble resources. Moreover, multiple handshakes during the access process result in high signaling overhead and high latency, especially in short packet transmission. 

Against the above background, grant-free random access has received significant attention in recent years, whereby MTDs can transmit data to evolved node base station (eNB) in an “arrive and go” manner. At the same time, non-orthogonal multiple access (NOMA) has been extensively studied for massive access in mMTC systems. The key idea of NOMA is to share the radio resources in a non-orthogonal manner in the power or code domain such that several times more users can be supported \cite{9369968}. Under this context, grant-free transmission using NOMA is considered as a promising solution \cite{7842433,9097306}.

\subsection{Related Works}
There are a number of proposals for grant-free transmissions \cite{8674774,9782529,10413949,10295166}. Firstly, due to the sporadic transmission in mMTC, the multi-user detection problem can be formulated as a sparse signal recovery problem, called compressive random access \cite{8674774,9782529}. Besides, spreading-based NOMA has also been studied for grant-free transmission. Yang $\textit{et al.}$ in \cite{10413949} proposed a two-stage active user detection including initial detection and false alarm correction for grant-free NOMA transmission, with which significantly improved active user detection and symbol error rate were achieved. For grant-free code-domain NOMA transmission, a network coding-assisted $K$-repetition scheme for sparse code multiple access was proposed in \cite{10295166}. The aforementioned grant-free NOMA transmission schemes support massive connectivity from the physical layer perspective. Moving up to the medium access control (MAC) layer, an important direction is to integrate slotted ALOHA with power-domain NOMA. For simplicity, the resultant system is called NOMA-ALOHA. Specifically, NOMA-ALOHA scheme was first considered in \cite{8085106}, in which the throughput of uplink multichannel NOMA-ALOHA was analyzed. The throughput of NOMA-ALOHA was studied in \cite{9410347} by considering two different arrival models including Binomial and Poisson arrivals. Moreover, the throughput of NOMA-ALOHA was studied with the assumption of perfect successive interference cancellation (SIC) \cite{9590503,9200648} and imperfect SIC \cite{10318204}. Recently, a novel received-power model was designed in \cite{10311394} to enhance the throughput of NOMA-ALOHA by improving SIC decoding performance, but at the cost of high transmit power. \cite{9521536} and \cite{10008538} introduced a power level selection scheme to improve the system throughput by minimizing the impact of user collisions on SIC decoding, aiming to obtain the maximum system throughput. Since higher throughput can reduce the access delay, the analysis of the age of information (AoI) over the uplink NOMA-ALOHA transmissions were studied in \cite{10311394,10260312}.

There are also some works concerning the energy efficiency of NOMA-ALOHA systems. To reduce the average transmit power, a channel-dependent (CD) energy efficient selection scheme was introduced in \cite{8085106} by setting the channel gain threshold for each power level. For the same purpose, a novel received-power model was designed in \cite{10311394} by ignoring the impact of multi-user interference on the SIC process, with which the average transmit power was reduced. However, these schemes may not be feasible when the network is overloaded. Several random access control schemes have been investigated for NOMA-ALOHA scheme \cite{9878108,9351544,9900347,9410347}. For a higher user load, the conventional ACB scheme \cite{9878108} and online backoff scheme \cite{9351544} were applied to NOMA-ALOHA, which can be seamlessly integrated into the LTE random access procedure. However, the ACB factor is sub-optimal in the sense that the optimal load of NOMA-ALOHA can be hardly obtained. To address this problem, \cite{9410347} and \cite{9900347} made an attempt to maximize the system throughput based on a throughput upper bound of NOMA-ALOHA.     

Recently, priority random access (PRA) has been studied in grant-based random access \cite{9174796,8422775,10171175}. These PRA schemes are typically built by assigning different access resources to different categories. A remarkable work was reported in \cite{9174796}, in which the access resources are divided into multiple groups corresponding to the priorities of different MTDs. An online control algorithm was then introduced to adaptively adjust the sizes of resource groups. \cite{8422775} and \cite{10171175} supported access priority by adaptively adjusting the ACB factors for MTDs with different priorities. In \cite{9539874}, the low-priority MTDs were restricted through the ACB scheme in order to create more access opportunities for high-priority MTDs. Moreover, the PRA scheme can be redesigned from a resource quality viewpoint \cite{9222516}, in which orthogonal preambles belonging to the high-priority MTDs enjoy a higher detection probability than that of the non-orthogonal preambles for the low-priority MTDs. A priority access (PA) scheme based on NOMA-ALOHA is proposed in \cite{9900347}, where each received power level corresponds to a specific device type for selection.

\subsection{Motivations \& Contributions}

Existing state-of-the-art NOMA-ALOHA schemes are primarily developed based on received power level models under ideal conditions, but their performance may be severely degraded in practical scenarios. In particular, imperfect channel state information (CSI) and imperfect SIC introduce residual interference, which may substantially degrade the signal decoding performance of NOMA-ALOHA systems. Consequently, these schemes may suffer from limited applicability in real-world networks. In contrast to existing studies, this work explicitly incorporates imperfect CSI and imperfect SIC into the received power level modeling of NOMA-ALOHA, enabling reliable system operation under non-ideal conditions.

Moreover, in heterogeneous mMTC scenarios, different types of MTDs may transmit simultaneously and exhibit heterogeneous QoS requirements, rendering existing non-prioritized NOMA-ALOHA schemes insufficient to meet diverse service demands \cite{8085106,9410347,9590503,9200648,10318204}. Although a PA scheme for NOMA-ALOHA has been proposed in \cite{9900347}, it is designed under relatively static assumptions and does not explicitly consider the impact of real-time traffic dynamics in practical networks. In realistic mMTC systems, variations in traffic load and in the proportion of MTDs across different priority classes may significantly affect the performance of PRA schemes. To address this issue, this work investigates PRA for NOMA-ALOHA under dynamic traffic conditions, aiming to support differentiated QoS requirements in practical mMTC systems.

In addition, NOMA-ALOHA is also affected by the user overload problem, leading to a significant deterioration in throughput performance. Although such a problem can be alleviated through random access control, it may cause high transmit power. For example, in the existing user barring algorithm (UBA) \cite{9410347}, the MTDs are required to generate a random number and compare it with the user barring factor to decide whether to transmit data or not. As a result, MTDs with high channel quality may be prohibited from transmitting, while MTDs with poor channel quality are allowed to transmit data. Since the channel inverse-based transmit power allocation is typically used in NOMA-ALOHA, the MTDs with poor channel quality may have to require high transmit power, which is unaffordable for MTDs with low power requirement. 

Motivated by the above considerations, we aim to achieve priority-based random access for NOMA-ALOHA in heterogeneous mMTC under non-ideal conditions, while effectively alleviating the user overload problem and the energy consumption. The main contributions are summarized as follows:

\begin{itemize}
\item We develop a received power level model for NOMA-ALOHA under imperfect CSI and SIC. The effective range of SIC efficiency is analyzed to ensure the stable operation of NOMA-ALOHA under non-ideal conditions. Leveraging this model, we design a PRA scheme for NOMA-ALOHA (PRA-NA), which provides access priority by allocating power levels with varying quantities and qualities to different types of MTDs. 

\item Considering that the number of different types of MTDs may change in real-life networks, two PRA-NA schemes including fixed PRA-NA (FPRA-NA) and adaptive PRA-NA (APRA-NA) are designed to reduce the average access delay of delay-sensitive MTDs. Moreover, the upper-bound throughputs are analyzed for both FPRA-NA and APRA-NA schemes, allowing us to find the optimal load that maximizes the system throughput.

\item To improve the energy efficiency and alleviate the user overload problem, an enhanced UBA (EUBA) is proposed in this paper. Our main innovation is to utilize channel quality threshold as a critical component of the barring factor to control the load of network. Firstly, the optimal channel quality threshold is analyzed to stabilize the system throughput. Then, the EUBA is extended to the proposed PRA-NA scheme to provide access priority, which in turn effectively alleviates the user overload problem and energy consumption.

\item Finally, extensive simulation experimental results are shown adopting the 3GPP traffic model following Beta distribution (3GPP TR 37.868) \cite{37868}. Our simulation results validate the effectiveness of the proposed schemes in terms of throughput, average transmit power, and average access delay.
\end{itemize}

The remainder of this paper is organized as follows. The system model and the proposed received power level model for NOMA-ALOHA are introduced in Section \ref{section 2}. The PRA-NA scheme is proposed and mathematically analyzed in Section \ref{section 3}. The EUBA is developed and extended to the proposed PRA-NA scheme in Section \ref{section 4}. Simulation results are presented in Section \ref{section 5}. Finally, this paper is concluded in Section \ref{section 6}.

$\mathit{Notations}$: In this paper, $\mathcal{CN}(\textbf{a},\textbf{R})$ represents the distribution of circularly symmetric complex Gaussian (CSCG) random vectors with mean vector \textbf{a} and covariance matrix \textbf{R}. $\mathbb{E}[ \cdot]$ denotes the statistical expectation. For a set $\mathcal{A}$, $|\mathcal{A}|$ denotes the number of elements in the set $\mathcal{A}$. The Beta function is given by $\boldsymbol{B}(\alpha,\ \beta)=\int_{0}^{1} x^{\alpha -1}(1-x)^{\beta -1}\mathrm{d}x, \ \alpha,\ \beta>0$.

\section{ System Model}\label{section 2}

In this paper, we consider a slotted-ALOHA uplink transmission network consisting of $M$ MTDs, communicating with the same eNB. Assume that each MAC frame comprises $S$ time slots, each MAC frame has a duration of $\tau$ seconds. Similar to \cite{8085106,9410347,9590503,9200648,10318204}, due to the channel reciprocity in the time division duplexing (TDD) mode, let us assume that the channel coefficient can be estimated at the active MTDs through the downlink transmission stage using the well-known pilot-based channel estimation technique.

NOMA-ALOHA can support the successful transmission of multiple packets in a slot with multiple power levels by performing SIC. At the start of each frame, the active MTD randomly select a time slot and a power level for uplink transmission, by assuming that each MTD is frame and slot synchronized. Denote by $N$ the number of active users in a slot. Then, the received signal at the eNB can be formulated as 
\begin{align}\vspace{-0.1cm}
y=\sum_{n=1}^{N}\left(\hat{h}_{n}+\epsilon _n\right)\sqrt{\rho_{n}}s_{n}+n_{0},  \label{e1}
\end{align}
where $\hat{h}_{n}$ represents the estimated channel coefficient between eNB and the $n$-th MTD, and  $\epsilon_n \sim \mathcal{CN}(0,\sigma_e^{2})$ is the channel estimation error with mean zero and variance $\sigma_e^{2}$. Similar to \cite{6400277}, the channel estimation error is modeled as additional Gaussian noise. $\rho_{n}$ and $s_{n}$ represent the transmit power and signal of MTD $n$, respectively. $n_{0} \sim \mathcal{CN}(0,\sigma^{2})$ denotes additive white Gaussian noise where $\sigma^{2}$ is the noise power. In this paper, the noise power is assumed to be normalized without loss of generality.
  
In order to implement SIC at the eNB, assume that $Q>1$ received power levels are pre-determined for active users to choose randomly, denoted by 
\begin{align}r_{1}>r_{2}>\dots >r_{q}>\dots>r_{Q}>0, \label{e2}\end{align}
where $r_ {q} $ represents the received power value at the eNB for a MTD that selects the $q$-th power level. Let $\Gamma$ denote the target signal-to-interference-plus-noise ratio (SINR), similar to \cite{8085106,9410347,9590503,9200648,10318204}, an equal target SINR is assumed for all MTDs to reduce the system complexity of the NOMA-ALOHA scheme. Considering imperfect SIC, similar to \cite{9573456}, we define $\delta \in (0,1)$ as the SIC efficiency \footnote{Although the SIC efficiency may depend on the decoding order and instantaneous interference conditions, it is commonly modeled as a constant to characterize the average residual interference under imperfect SIC, enabling a tractable system-level analysis.}. Accordingly, the received power level model is derived as
  \begin{align}r_{q}=\frac{\sigma^2\Gamma \gamma^{Q-q}(\gamma-1)}{(1-\Gamma\sigma_e^2)(\gamma-1)-\Gamma \delta \left(\gamma^{Q}-\gamma\right)},  \label{e3}\end{align}
where $\gamma=\frac{1+(1-\sigma_e^2)\Gamma}{1+(\delta-\sigma_e^2)\Gamma}$, and the SIC efficiency is required to satisfy $\delta\le\delta^{*}$, where $\delta^{*} =\min\left(\frac{(Q-1)(1-\sigma_e^2)\Gamma-1}{Q\Gamma}, \frac{1}{\left(1+\Gamma/\left(1-\Gamma\sigma_e^2\right)\right)^{Q}-\Gamma/\left(1-\Gamma\sigma_e^2\right)-1}\right)$. The proof is provided in Appendix A.

The received power values of all power levels can be stored in MTDs. When an MTD $n$ attempts to transmit data, it randomly selects a power level $q$, and the corresponding received power value is then determined accordingly.  For example, let $Q = 4$,  $\Gamma=2$, $\sigma^2=1$, $\sigma_e^2=10^{-4}$, and $\delta=\delta^*\approx\ 0.0128$. According to \eqref{e3}, we have 
\begin{align}
   r_1\approx783.34, & \ r_2\approx267.77, \nonumber\\ 
 r_3\approx91.63,& \ r_4\approx31.28. \label{e4}
\end{align}
Without loss of generality, the SIC decoding is performed in descending order of the received power levels, i.e., the signal corresponding to $r_1$ is decoded first, followed by $r_2,\ldots,r_Q$. Based on the received power model, if only one active user exists at each power level, the SINR for the active MTD that chooses $r_1$ becomes $\frac{r_1}{\sum_{i=2}^{Q}r_i+r_1\sigma_e^2+\sigma^2}$,
which equals $\Gamma$. Similarly, the SINR for the active MTD choosing $r_2$ becomes $\frac{r_2}{\sum_{i=3}^{Q}r_i+\sum_{i=1}^{1}\delta r_i+r_2\sigma_e^2+\sigma^2}$, which also equals $\Gamma$. Hence, similar to \cite{8085106,9410347,9590503,9200648,10318204}, all the $Q$ signals can be decoded using SIC. However, when $\delta> \delta^{*}$, the received power model becomes infeasible. For example, when $\delta=0.02$, we have  
\begin{align}
   r_1\approx-177.41, & \ r_2\approx-40.69, \nonumber\\ 
 r_3\approx-14.10,& \ r_4\approx-4.88, \label{e5}
\end{align}
the insufficient SIC efficiency results in strong residual interference, thereby preventing the generation of viable received power levels under the proposed received power level model.

\begin{figure}[t!]\centering \vspace{-0em}
	\epsfig{file=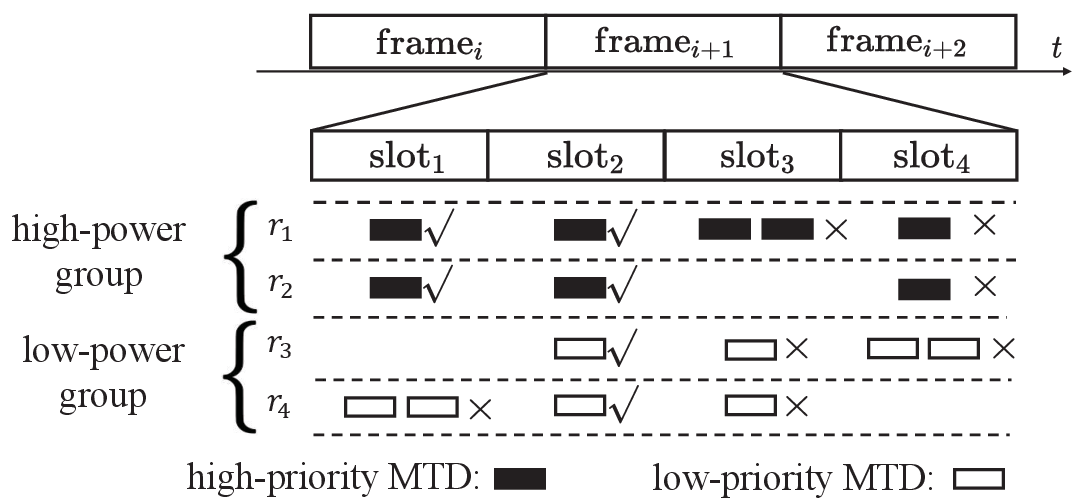, width=0.45\textwidth, clip=}\vspace{-0.5em}
	\caption{ The proposed PRA-NA scheme. 
		\vspace{-1em}    }\label{fig1}   \vspace{-1.5em} 
\end{figure}
\section{PRA scheme for NOMA-ALOHA Uplink Transmission}  \label{section 3}

In NOMA-ALOHA, if two or more active MTDs select the same power level in a slot, power collision occurs, resulting in transmission failure for all active MTDs in this power level. Moreover, signals with decoding failure at power level $q$ are considered as the interference at power levels $q^{\prime}>q$. Thus, the power collisions in different power levels have different impacts on decoding in other power levels. For example, let system parameters $Q=4$, $\Gamma=2$, $\sigma^2=1$, $\sigma_e^2=10^{-4}$, and $\delta=0.001$. From \eqref{e3}, the received power values of all power levels can be obtained as $\left \{ r_{1},r_{2},r_{3},r_{4} \right \} = \left \{ 58.23,19.44,6.49,2.16 \right \}$. When power collisions occurs at the power levels with higher received power (e.g., $r_{1}$ and $r_{2}$), let us assume $\left \{ N_{1}, N_{2}, N_{3}, N_{4} \right \} =\left \{ 2,1,1,0 \right \}$, where $N_{q}$ is defined as the number of active MTDs choosing the $q$-th power level. Since power collision occurs at the power level 1, the signals at the power level 1 are considered as the interference at the power levels $q>1$. Then, the SINRs for the active MTDs choosing $r_{2}$ and $r_{3}$ are $\frac{r_{2}}{2r_{1}+r_{3}+r_2\sigma_e^2+1}\approx\frac{19.44}{123.95}\approx 0.15 $ and $\frac{r_{3}}{2r_{1}+r_{2}+r_3\sigma_e^2+1}\approx\frac{6.49}{136.90}\approx 0.04 $, which are less than $\Gamma=2$. It is obvious that when power collision occurs at power level $q$, decoding failure occurs at power levels $q^{\prime}>q$ due to error propagation in SIC. However, when power collisions occur at the power levels with lower received power (e.g., $r_{3}$ and $r_{4}$), it leads to two different consequences on the decoding at the power levels with higher received power. For example, when $\left \{ N_1,N_2,N_{3}, N_{4} \right \} =\left \{ 1,1,2,0 \right \}$, the SINRs for the active MTDs choosing $r_{1}$ and $r_{2}$ are $\frac{r_{1}}{r_{2}+2r_{3}+r_1\sigma_e^2+1}\approx\frac{58.23}{33.44}\approx 1.74$ and $\frac{r_{2}}{r_{1}+2r_{3}+r_2\sigma_e^2+1}\approx\frac{19.44}{72.22}\approx 0.26$, which are less than $\Gamma=2$. Thus, one consequence is the decoding failure at the power levels with higher received power. However, when $\left \{ N_1,N_2,N_{3}, N_{4} \right \} =\left \{ 1,1,0,2 \right \}$, the SINRs for the active MTDs choosing $r_{1}$ and $r_{2}$ are $\frac{r_{1}}{r_{2}+2r_{4}+r_1\sigma_e^2+1}\approx\frac{58.23}{24.78}\approx 2.34$ and $\frac{r_{2}}{2r_{4}+\delta r_1+r_2\sigma_e^2+1}\approx \frac{19.44}{5.39}\approx 3.62$, which are greater than $\Gamma$. In this case, the signals at the power levels with higher received power can be successfully decoded. Since all active MTDs independently choose their power levels at the identical distribution (i.e., uniform distribution), the probability of a power collision occurring at each power level is same. Accordingly, the transmissions in power levels with higher received power might be more reliable than power levels with lower received power. 

Based on above observation, a PRA-NA scheme is proposed to provide access priority as shown in Fig. 1. Firstly, similar to \cite{8422775,9222516}, the MTDs in heterogeneous mMTC can be classified into two categories: delay-sensitive MTDs and delay-tolerant MTDs. Delay-sensitive MTDs (e.g., sensors in vehicle to everything and public safety devices) require a high successful transmission probability and short delay requirements. Thus, the delay-sensitive MTDs can be seen as high-priority MTDs. For delay-tolerant MTDs (e.g., smart meters, etc.), the MTDs report data periodically to the eNB with less stringent access delay constraints (e.g., a half hour). Thus, the delay-tolerant MTDs can be seen as low-priority MTDs. Then, the power levels are divided into two groups including high-power group (e.g., $r_{1}$ and $r_{2}$) and low-power group (e.g., $r_{3}$ and $r_{4}$). The high-power group serves the high-priority MTDs, and low-power group serves the low-priority MTDs, respectively. Due to the fact that the transmissions in high-power group are more reliable than low-power group, the re-transmission probability can be reduced for high-priority MTDs, thereby the improvement of access delay performance for high-priority MTDs in the proposed scheme can be achieved. The throughput analysis of the proposed PRA-NA scheme is presented in Subsection~\ref{section 3a}. The corresponding notations for the proposed PRA-NA scheme are explained in Table~\ref{tab:tab1}.

\subsection{Throughput Analysis of the PRA-NA Scheme}\label{section 3a}

\begin{table}[t!]
	\centering
	\caption{Summary of Main Notations }\label{tab:tab1}
	\begin{tabular}{p{1.5cm}  p{6cm}  }
		\hline
		$\textbf{Notation}$  &  $\textbf{Explanation}$ \\
		$Q$ & the number of power levels\\
		$N$ &  the number of active MTDs in a time slot \\
        $\mu$& the proportion of high-priority MTDs\\ 
        $Q_{1}$  & the number of power levels in high-power group \\
		$Q_{2}$  & the number of power levels in low-power group \\
        $\eta$& the throughput of the PRA-NA scheme\\
        $\eta_{h}$& the average number of high-priority MTDs successful access in the PRA-NA scheme\\
        $\eta_{l}$& the average number of low-priority MTDs successful access in the FPRA-NA scheme\\
        $N_{i}$&  the number of active MTDs in the $i$-th power level\\

		\hline
	\end{tabular}
\end{table}

In this paper, the throughput is defined as the average number of MTDs with successful access per slot (i.e., the packets can be successfully decoded at the eNB). As previously analyzed, when power collisions occur at the power levels with lower received power, the decoding at the power levels with higher received power may yield two different consequences (i.e, decoding successful and decoding failure). As $N$ and $Q$ increase, it is difficult to analyze the probability of the packet with successful decoding at each power level, making it challenging to derive an expression for the exact throughput. Consequently, the exact optimal load can be hardly analyzed to maximize system throughput. To address this issue, similar to \cite{9410347,9590503,10318204}, we assume that power collisions at the power levels with lower received power do not affect decoding at the power levels with higher received power. In this case,  the throughput derived can be served as an upper bound. Although this assumption is difficult to achieve in reality, the optimal load analyzed based on the throughput bound has been proven effective in existing researches \cite{9410347,9590503,10318204}. We consider a single slot with $Q$ power levels, where the average number of active MTDs is $N$ and the proportion of high-priority MTDs is $\mu$. The numbers of power levels in the high-power and low-power groups are denoted by $Q_1$ and $Q_2$, respectively, with $Q_1+Q_2=Q$. Let $\eta$ denote the throughput of the PRA-NA scheme. Then, we have
\begin{align}\eta =\eta_{h}+\eta_{l}, \label{e6}\end{align}
where $\eta_{h}$ and $\eta_{l}$ represent the average numbers of MTDs with successful access in the high-power and low-power groups, respectively.

For the high-power group, similar to \cite{9410347,9590503}, we assume that the number of high-priority MTDs selecting each power level follows a Poisson distribution with mean $\mu N/Q_1$. Thus, we have
\begin{align}
a &= P(N_q=0)=e^{-\frac{\mu N}{Q_1}}, \nonumber\\
b &= P(N_q=1)=\frac{\mu N}{Q_1}e^{-\frac{\mu N}{Q_1}}. \label{e7}
\end{align}
The MTD at the $q$-th power level can be successfully accessed only when exactly one high-priority MTD selects this power level and no power collision occurs at any power level $q^{\prime}<q$. Therefore, the average number of successfully accessed high-priority MTDs is given by
\begin{align}
\eta_h
&= \sum_{q=1}^{Q_1} P(N_q=1)\prod_{i=1}^{q-1}P(N_i\le 1) \nonumber\\
&= \sum_{q=1}^{Q_1} b(a+b)^{q-1} \nonumber\\
&= b\frac{1-(a+b)^{Q_1}}{1-(a+b)}.
\label{e8}
\end{align}

For the low-power group, let $c$ and $d$ be the probabilities that $N_p=0$ and $N_p=1$ in the low-power group, respectively. Similarly, assuming that the number of low-priority MTDs selecting each power level follows a Poisson distribution with mean $(1-\mu)N/Q_2$, we have \begin{align}& c=P\left(N_{p}=0\right)=e^{-\frac{(1-\mu) N}{Q_2}}\nonumber
\\ &d=P(N_{p}=1)=\frac{(1-\mu)N}{Q_2} e^{-\frac{(1-\mu) N}{Q_2}}. \label{e9}
\end{align}
Similarly, the MTD at the $p$-th power level can be successfully accessed only when exactly one low-priority MTD selects this power level, no power collision occurs at any power level in the high-power group, and no power collision occurs at any power level $p^{\prime}<p$ in the low-power group. Therefore, the average number of successfully accessed low-priority MTDs is given by
\begin{align}
\eta_l
&= (a+b)^{Q_1}\sum_{p=1}^{Q_2} P(N_p=1)\prod_{i=1}^{p-1}P(N_i\le 1) \nonumber\\
&= (a+b)^{Q_1}\sum_{q=1}^{Q_2} d(c+d)^{p-1} \nonumber\\
&= d(a+b)^{Q_1}\frac{1-(c+d)^{Q_2}}{1-(c+d)}.
\label{e10}
\end{align}

\subsection{FPRA-NA and APRA-NA }\label{section 3b}
Since the numbers of high-priority and low-priority MTDs may vary in practical networks, FPRA-NA and APRA-NA are designed for different traffic scenarios.

Considering the scenario where there are fewer high-priority MTDs in the network, the traffic fluctuation of the high-priority MTDs in the network is small (i.e., the variation in the number of high-priority MTDs across different frames is small). Consequently, the fluctuation in the number of access resources required for high-priority MTDs is also small. Motivated by this observation, an FPRA-NA scheme is designed by keeping the numbers of power levels in the high-power and low-power groups fixed in each uplink transmission frame. Let $F_h$ and $F_l$ denote the average numbers of successfully accessed high-priority and low-priority MTDs in the proposed FPRA-NA scheme, respectively. We have
\begin{align}
F_h=\eta_h,\ \ F_l=\eta_l.
\label{e15}
\end{align}
Let $\eta_f$ denote the throughput of FPRA-NA scheme, which can be written as
\begin{align}
\eta_f=F_h+F_l,
\label{e16}
\end{align}
where $Q_1$ and $Q_2$ are fixed in each uplink transmission frame.

When the high-priority MTDs experience high traffic, the traffic fluctuation of the high-priority MTDs in the network may be significant (i.e., large variations in the number of high-priority MTDs across different frames), implying that the number of access resources required for high-priority MTDs also fluctuates significantly. Therefore, an APRA-NA scheme is designed to reduce the access delay of high-priority MTDs by dynamically adjusting the number of power levels allocated to the high-power group in each uplink transmission frame.

In the proposed APRA-NA scheme, similar to \cite{9174796}, the average resource occupancy of different priority MTDs is proportional to the priority weight. We assign priority weights $w_{h}$ and $w_{l}$ for high-priority MTDs and low-priority MTDs, respectively, where $w_{h}/w_{l}=w $. Since the proportion of high-priority MTDs is $\mu$, we have
\begin{align}\frac{Q_{1}}{Q_{2}} 
	=  \frac{\mu w}{1-\mu } \nonumber \\
	Q_{1}+Q_{2} = Q.  \label{e17}
\end{align}
Accordingly, we obtain $Q_{1}=\frac{\mu wQ}{1+(w-1)\mu}$ and $Q_{2}=\frac{(1-\mu)Q}{1+(w-1)\mu}$. 

Let $A_h$ and $A_l$ denote the average numbers of successfully accessed high-priority and low-priority MTDs in the proposed APRA-NA scheme, respectively. Then, we have 
\begin{align} 
A_{h}=\eta_{h},\ \  A_{l}=\eta_{l}. 
\label{e18}
\end{align}
Let $\eta_{a}$ denote the throughput of the APRA-NA scheme, which is given by
\begin{align}
\eta_{a} =A_{h}+A_{l},  
\label{e19}
\end{align}
where $Q_1$ and $Q_2$ are dynamically adjusted according to the proportion of high-priority MTDs in each uplink transmission frame. Note that FPRA-NA can be regarded as a special case of APRA-NA when the proportion of high-priority MTDs in each slot remains unchanged and the high-priority weight is fixed.

\section{Enhanced User Barring Algorithm for PRA-NA Scheme}\label{section 4}

Similar to the conventional NOMA-ALOHA scheme, the proposed PRA-NA also suffers from the user overload problem, which leads to significant degradation in throughput and access delay performance. To mitigate this issue, various random access control schemes, such as user barring algorithms (UBA) \cite{9410347}, access class barring (ACB) \cite{9539874}, and transmission probability control \cite{10260312}, have been widely adopted to regulate the access attempts of MTDs and stabilize the system throughput of NOMA-ALOHA schemes. In these schemes, before transmission, each MTD generates a random number and compares it with the received barring factor. If the generated random number is smaller than the barring factor, the MTD is allowed to transmit; otherwise, the access attempt is deferred. However, since channel inversion-based power control is typically employed in NOMA-ALOHA, MTDs with poor channel quality may be required to transmit at excessively high power levels when selecting higher received power levels, resulting in low energy efficiency.

To reduce the average transmit power, a natural approach is to restrict access attempts from MTDs with poor channel quality, which can be viewed as a channel quality-based UBA. However, such a strategy may introduce location-dependent unfairness. For instance, MTDs located far from the eNB may rarely obtain transmission opportunities, even when carrying high-priority data. To strike a balance between energy efficiency and location-dependent fairness, we propose an enhanced user barring algorithm (EUBA) that integrates conventional UBA with channel quality-based UBA. Specifically, an active MTD first performs conventional UBA. If it passes, the MTD is allowed to transmit immediately. Otherwise, it further undergoes channel quality-based UBA. If this second criterion is satisfied, the MTD transmits; otherwise, it waits for a subsequent transmission opportunity. In the proposed EUBA, conventional UBA helps preserve fairness across different locations, while channel quality-based UBA effectively reduces the average transmit power.

\subsection{EUBA } \label{section 4a}
In this paper, denote $\tau_d$ the average access delay. Since higher throughput leads to lower average number of retransmissions, the access delay minimization problem can be formulated as a throughput maximization problem: 
\begin{align}
	\arg  \min \ \tau_d=\arg \max \ \eta,  \label{e20}
\end{align}
the number of active MTDs for each slot can be obtained by setting $\frac{\partial \eta} {\partial N}$ to zero, i.e., $N_{*}$. Denote $\lambda_{*}$ the optimal number of active MTDs for each frame, which can be calculated by $\lambda_{*}=SN_{*}$.

The UBA can be viewed as a Bernoulli trial
where the probability of success is the barring factor $P_b$. Define $\lambda$ as the number of active MTDs during a frame before performing UBA. The number of active MTDs after UBA execution $\xi$ follows a binomial distribution with parameter $P_{b}$, i.e., $\xi \sim \mathbf{B} \left (  \lambda, P_{b}\right )$.  To achieve maximum throughput, the average number of active users after UBA should be maintained at the optimal number (i.e., $\lambda_*$). Thus, we have 
\begin{align}
	\mathbb{E}[\xi ]=\lambda _{*}=\lambda P_{b}.  \label{e21}
\end{align}
Accordingly, the optimal barring factor can be obtained by $P_{b}^{*}=\min \left(\frac{\lambda_{*}}{\lambda},1\right)$.
To achieve a trade-off between average transmit power and location-dependent fairness, the proposed EUBA employs a trade-off parameter $\theta\in (0,1)$, and the barring factor can be expressed as $P_b=\theta P_b^*$. Thus, the active MTDs first perform UBA with the barring factor $\theta P_b^*$. Then, channel quality-based UBA is introduced to barring the remaining active MTDs with the channel quality threshold, which can reduce the average transmit power. 

To obtain the optimal channel quality threshold, similar to \cite{8085106}, we assume that the active MTDs are uniformly distributed within a cell of radius $R$, and the channel coefficients between MTDs and eNB could be known at MTDs. A standard power-law path-loss model is adopted to model the channel. The channel coefficient can be written as $h_n=u_nd_n^{-\kappa/2}$, where $u_n$ represents the small-scale fading coefficient and is an independent Rayleigh random variable with $\mathbb{E} [u_n^2 ] = 1$. $d_{n}$ represents the distance between the MTD $n$ and eNB, $\kappa$ the path loss exponent. Thus, the channel quality could be expressed as 
\begin{align}
	\mathbb{E}[\left | h_{n} \right |^{2} ]=d_{n}^{-\kappa },  \ \ \   0<d_{n} \le R    \label{e22}
\end{align} 
Denote by $H^{*}$ the optimal channel quality threshold, we have
\begin{align}
H^{*}=d_{*}^{-\kappa },   \label{e23}
\end{align} 
where $d_{*}$ is the optimal distance. The active MTDs that satisfy the channel quality threshold $H^*$ are allowed to transmit and form a group
\begin{align}
	\mathcal{A}=&\{n\ | \ d_{n}\le d_{*} \} .  \label{e24}
\end{align}
where $|\mathcal{A}|$ is the number of the MTDs that are permitted to transmit data in a frame. After performing UBA with barring factor $\theta P_b^*$, an average of $\theta \lambda_*$ MTDs are allowed to transmit. To achieve the optimal total of $\lambda^*$ active MTDs allowed to transmit, channel quality-based UBA should permit an average of $(1-\theta )\lambda^* $ MTDs to transmit from the remaining active MTDs after UBA, i.e., $\lambda - \theta \lambda^*$ active MTDs. Since MTDs are uniformly distributed within a cell, the optimal distance can be written as
\begin{align}
d_{*}=R\sqrt{\frac{(1-\theta)\lambda_{*} }{(1-\theta P_b^*)\lambda} } .   \label{e25}
\end{align}
To stabilize the NOMA-ALOHA system, the optimal load derived by lower-bound throughput is used in this paper\footnote{Here, we use the optimal load derived by the lower-bound throughput to analyze the optimal channel quality threshold. Because it is difficult to derive a closed expression of optimal load based on the upper-bound throughput. In addition, the optimal load based on the upper-bound throughput may unstabilize the NOMA-ALOHA system, because the load may be increased to meet the throughput based on an upper-bound throughput, which could lead to system overload. In this sense, the optimal load based on lower-bound throughput could be useful.}, where $N_*=\sqrt{Q}$ for a single slot with $Q$ power levels \cite{9590503}. As there are $S$ slots in a frame in this paper, the optimal load $\lambda_{*}$ can be expressed as $S\sqrt{Q}$ in a frame. 

To calculate the optimal distance, similar to \cite{9410347}, we assume that the idle power levels can be observed during the decoding, and the number of active MTDs in the next frame can be approximately estimated by the current load  $\tilde{\lambda}$. Denote $\tilde{N}_{s}$ as the estimated number of active MTDs transmitted in the $s$-th slot, we have $\tilde{\lambda}_{*}=\sum_{s=1}^{S}\tilde{N}_{s} $, where $\tilde{\lambda}_{*}$ is the estimated number of active MTDs  transmit data at the current frame. Similar to \eqref{e7} and \eqref{e9}, the average number of idle power levels in the $s$-th slot $Q_{idle}^{s}$ can be expressed as
\begin{align} 
\mathbb{E}[Q_{idle}^{s}] =Q e^{-\frac{\tilde{N}_{s}}{Q}}  \label{e26}
  \end{align}
where $Q_{idle}^{s}$ decreases monotonically as $\tilde{N}_{s}$ increases. Thus,  $\tilde{N}_{s}$ can be calculated according to the observed $Q_{idle}^{s}$. Therefore,  the number of active MTDs in the next frame can be approximately estimated by
\begin{align} 
\lambda\approx \tilde{\lambda}=\left(\frac {R}{\tilde{d}_{*}}\right) ^{2}\sum_{s=1}^{S}\tilde{N}_{s},   \label{e27}
\end{align}
where $\tilde{d}_{*}$ is the current optimal distance. According to the estimated load $\lambda$ and optimal load $\lambda_{*}=S\sqrt{Q}$, the optimal channel quality threshold can be calculated based on \eqref{e25} and \eqref{e23}. Before the uplink transmissions, the active MTDs receive system information that includes the user barring factor and the channel quality threshold, which are used to perform UBA and channel quality-based UBA. Therefore, the proposed EUBA could mitigate the user overload problem by restricting the transmissions of MTDs requiring high transmit power. 

\begin{figure*}[t!]\centering \vspace{-0em}
	\epsfig{file=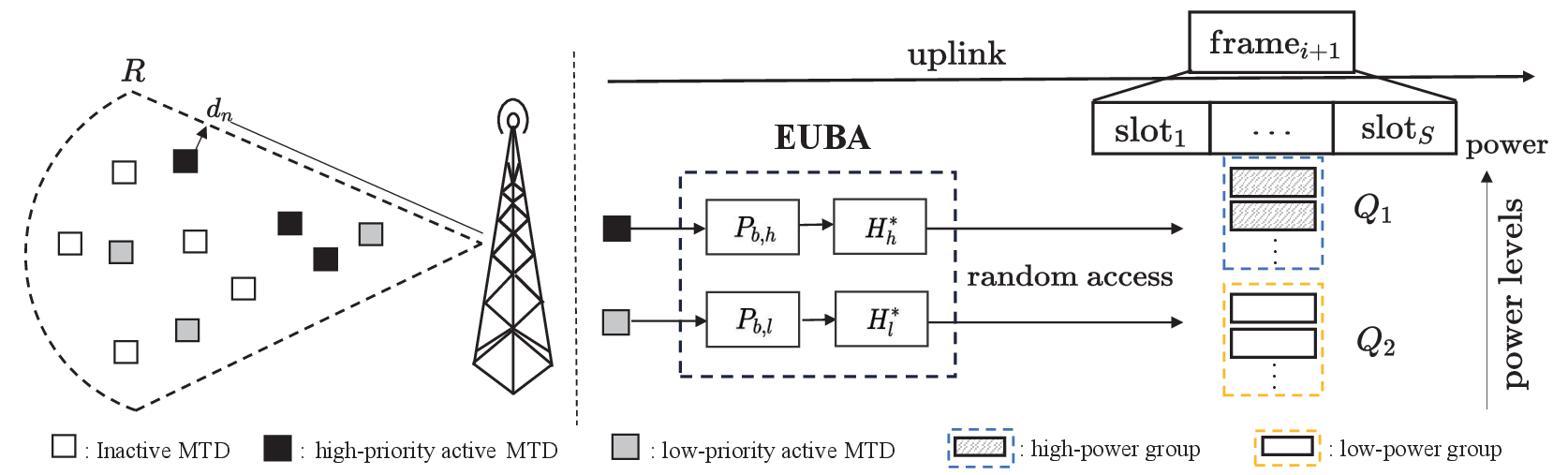, width=0.9\textwidth, clip=}\vspace{-0.5em}
	\caption{ The proposed PRA-NA scheme with EUBA. 
		\vspace{-1em}    }\label{fig2}   \vspace{-1.5em} 
\end{figure*}

\begin{algorithm}[t!]
	\caption{FPRA-NA with EUBA}
	\label{alg:Framwork}
	\KwIn{$S$, $T$, $Q$, $R$.}
	Initialize the number of power levels in high-power group and low-power group (i.e., $Q_{1}$, $Q_{2}$). Initialize the user barring factor $(P_{b,h}, \ P_{b,l})$ and the optimal channel quality threshold $(H^{*}_{h}, \ H^{*}_{l})$ for different priority MTDs. \\
	\For{ $t = 1,2,...,T$}{
		Perform EUBA: The active MTDs with different priorities generate a random number $p\in(0,1)$ and compare it with the corresponding user barring factor $(P_{b, \ h}, \ P_{b,l})$. Then, under imperfect CSI, the active MTDs with $p>P_{b}$ compare their estimated channel quality with the corresponding channel quality threshold $(H^{*}_{h}, \ H^{*}_{l})$;  \\
		The active MTDs with $p<P_{b}$ or $h_n>H^*$ are allowed to transmit data, and then randomly choose a slot and select a power level generated by \eqref{e3} from the corresponding power level group;\\
		\For{ $s = 1,2,...,S$}{
			Perform SIC with imperfect CSI and imperfect SIC at the receiver\\
			\eIf{power level $< Q_1$   }
			{
				Count the number of idle power levels in the $s$-th slot for high-power group, i.e., $Q_{idle}^{s,h}$;\\
			}
			{
				Count the number of idle power levels in the $s$-th slot for low-power group, i.e., $Q_{idle}^{s,l}$ ;\\
			}
			Calculate the number of different priority MTDs transmitted in the $s$-th slot according to \eqref{e26}, i.e., $\tilde{N}_{s,h},\tilde{N}_{s,l} $; \\
		}
		Calculate the current load according to \eqref{e27}, i.e., $\tilde{\lambda}_{h}, \tilde{\lambda}_{l}$;\\

        Update the user barring factor, i.e., $P_{b,\ h}, P_{b,\ l}$;\\
		
		Update the optimal distance according to \eqref{e25}, i.e., $d_{*,h}, d_{*,l}$;\\
		
		Update the optimal channel quality threshold according to \eqref{e23}, i.e., $H^{*}_{h}, H^{*}_{l}$;\\
		
	}
	Return the barring factor $(P_{b,\ h}, \ P_{b,\ l})$ and optimal channel quality threshold $(H^{*}_{h}, \ H^{*}_{l})$ for next frame traffic arrival. 
\end{algorithm}

\subsection{ FPRA-NA with EUBA } \label{section 4b}

In this subsection, we consider a small number of high-priority MTDs in the network. The FPRA-NA with EUBA under the heterogeneous mMTC scenarios is detailed in \textbf{Algorithm 1}. In the system setup, we consider $T$ uplink transmission frames, each of which consists of $S$ slots. The $Q$ received power levels are generated according to \eqref{e3}, while imperfect CSI and imperfect SIC are taken into account in the channel-quality estimation and SIC process. Step 1 in \textbf{Algorithm 1}, establishes the initial setting. Steps 2-4 of \textbf{Algorithm 1} show that MTDs perform the FPRA-NA scheme with EUBA, which can be shown in Fig. 2 where $Q_{1}$ and $Q_{2}$ are fixed. The user barring factor and the optimal channel quality threshold jointly constrain the number of active users to alleviate the user overload problem. The power levels are divided into two groups (i.e., high-power group and low-power group) to provide access priority. Steps 5-18 introduce the update process of the user barring factor and the optimal channel quality threshold for the next frame. According to the observed idle power levels $Q_{idle}^{s,h}$ and $Q_{idle}^{s,l}$ in Steps 5-11, the number of active transmitted data in the $s$-th slot (i.e., $\tilde{N}_{s,h}$ and $\tilde{N}_{s,l}$) and the estimated current load (i.e., $\tilde{\lambda}_{h}$ and $\tilde{\lambda}_{l}$) can be calculated in Steps 12-14. Finally, the barring factor (i.e., $P_{b,\ h}$ and $P_{b,\ l}$ ) and the optimal channel quality threshold (i.e., $H^{*}_{h}$ and $H^{*}_{l}$) for the next frame can be updated at the eNB.

\begin{algorithm}[b!]
	\caption{APRA-NA with EUBA}
	\label{alg:2}
	\KwIn{$S$, $T$, $Q$, $R$, $w$.}
	Initialize the number of power levels in high-power group and low-power group (i.e., $Q_{1}$, $Q_{2}$). Initialize the user barring factor $(P_{b,h}, \ P_{b,l})$ and the optimal channel quality threshold $(H^{*}_{h}, \ H^{*}_{l})$ for different priority MTDs. \\
	\For{ $t = 1,2,...,T$}{
		Perform EUBA: Similar to \textbf{Algorithm 1}; \\
		The active MTDs with $p<P_{b}$ or $h_n>H^*$ are allowed to transmit data, and then randomly choose a slot and select a power level generated by \eqref{e3} from the corresponding power level group;\\
		\For{ $s = 1,2,...,S$}{
			Perform SIC with imperfect CSI and imperfect SIC at the receiver\\
			\eIf{power level $< Q_{1}$   }
			{
				Similar to \textbf{Algorithm 1}, count $Q_{idle}^{s,h}$;\\
			}
			{
				Similar to \textbf{Algorithm 1}, count $Q_{idle}^{s,l}$ ;\\
			}
			Similar to \textbf{Algorithm 1}, calculate$\tilde{N}_{s,h},\tilde{N}_{s,l} $; \\
		}
		Calculate the current load according to \eqref{e27}, i.e., $\tilde{\lambda}_{h}, \tilde{\lambda}_{l}$;\\
		
		Calculate the proportion of high-priority active MTDs $\mu$, i.e., $\tilde{\lambda}_{h}/(\tilde{\lambda}_{h}+\tilde{\lambda}_{l}) $;\\
		
		Update the number of power levels for different priority MTDs according to \eqref{e17}, i.e., $Q_{1}, \ Q_{2}$\\

        Update the user barring factor, i.e., $P_{b,\ h}, P_{b,\ l}$;\\
		
		Update the optimal distance according to \eqref{e25}, i.e., $d_{*,h}, d_{*,l}$;\\
		
		Update the optimal channel quality threshold according to \eqref{e23}, i.e., $H^{*}_{h}, H^{*}_{l}$;\\
		
	}
	Return ($Q_{1}$, $Q_{2}$), the barring factor $(P_{b,\ h}, \ P_{b,\ l})$, and the optimal channel quality threshold $(H^{*}_{h}, \ H^{*}_{l})$ for next frame traffic arrival. 
\end{algorithm}

\subsection{APRA-NA with EUBA } \label{section 4c}

When high-priority MTDs experience high traffic, the proposed \textbf{Algorithm 1} cannot provide efficient access for high-priority MTDs since the traffic fluctuation of high-priority MTDs may be significant. Thus, in this subsection, APRA-NA is combined with EUBA to improve the access delay performance for high-priority MTDs while effectively improving energy efficiency and concurrently mitigating the user overload problem.

In \textbf{Algorithm 2}, the APRA-NA with EUBA is described under the heterogeneous mMTC scenarios. The system setup of \textbf{Algorithm 2} is consistent with that of \textbf{Algorithm 1}. In addition, the priority weights (i.e., $w_{h}$ and $w_{l}$) are preconfigured at the eNB, thus the parameter $w$ can be obtained. In Step 1, different from \textbf{Algorithm 1}, the number of power levels in high-power group and low-power group (i.e., $Q_{1}$ and $Q_{2}$) are initialized. Steps 2-4 of \textbf{Algorithm 2} can be shown in Fig. 2, which depicts that different MTDs perform EUBA under the APRA-NA scheme. In \textbf{Algorithm 2}, the number of power levels in high-power group $Q_{1}$ gradually increases as the number of high-priority MTDs rises, to accommodate the traffic fluctuations of the high-priority MTDs. Similar to \textbf{Algorithm 1}, the idle power levels $Q_{idle}^{s,h}$ and $Q_{idle}^{s,l}$ can be observed in Steps 5-11. Steps 12-20 are introduced to update parameters for the next frame, including the numbers of power levels in high-power group and low-power group, user barring factor, and optimal channel quality threshold for different priority MTDs.

The time complexity analysis of the proposed \textbf{Algorithm~1} and \textbf{Algorithm~2} is summarized as follows. 
For a single frame, the overall time complexity is dominated by the operations at both the MTD side and the eNB side. 
At the MTD side, $\lambda$ active MTDs perform EUBA in each frame, which incurs a computational complexity of $\mathcal{O}(\lambda)$. 
In addition, the selection of time slots and power levels for an average of $\lambda_*$ active MTDs requires $\mathcal{O}(\lambda_*)$, where $\mathcal{O}(\lambda_*) \le \mathcal{O}(\lambda)$. 
At the eNB side, the slot-level processing over $S$ time slots involves SIC decoding with a complexity of $\mathcal{O}(Q)$, idle power-level counting with $\mathcal{O}(Q)$, and the calculation of $\tilde{N}_{s,h}$ and $\tilde{N}_{s,l}$ with $\mathcal{O}(1)$ complexity per slot, resulting in a total complexity of $\mathcal{O}(SQ)$ per frame. 
Moreover, the parameter updates in each frame require $\mathcal{O}(1)$ complexity for both algorithms. 
Therefore, the single-frame time complexity of \textbf{Algorithm~1} and \textbf{Algorithm~2} is $\mathcal{O}(\lambda + SQ)$, and the overall time complexity over $T$ frames is $\mathcal{O}\!\left(T(\lambda + SQ)\right)$.

\begin{table}[t]
	\centering
	\caption{Simulation Parameters }\label{tab:tab2}
	\begin{tabular}{p{4.5cm}  p{3cm}  }
		\hline
		$\textbf{Parameter}$  &  $\textbf{Value}$ \\
		\hline
		Number of power level $Q$&  [2, 10]\\ 
		Number of slot  $S$&  8\\
		The traffic time $T_t$ & 1 s\\
		Total number of MTDs $M$ & $4000$\\
		MAC frame $\tau$ & 5 ms\\
        The number of  frame $T$ & 200\\
		Beta traffic model & $\alpha=3$ and $\beta=4$ \\
		Target SINR $\Gamma$ & [0, 10] dB \\
		Received-power levels & Equation \eqref{e3} \\
		& $\forall q=1,\dots,Q$\\
        SIC efficiency  $\delta$&  $0.1 \times \delta^{*}$ \\
        channel estimation error $\sigma_e^2$ & $10^{-6}$ \\
		Transmit power model  & Channel inversion \\
		Proportion of high-priority MTDs $\mu$  & [0, 1) \\
		Ratio of priority weight  $w$  & 1, 2 \\
		Path loss exponent $\kappa$ & 3.5 \\

		\hline
	\end{tabular}
\end{table}
\section{Simulation Results and Discussions}\label{section 5}

In this section, we present simulation results to compare the performance of the proposed scheme with the existing NOMA-ALOHA scheme in terms of throughput, average transmit power, and average access delay. The simulation parameters are listed in Table II. Moreover, to compare fairly with existing schemes, similar to \cite{8085106}, we assume that $R = 1$ for normalization purposes.

Moreover, a typical Beta traffic model proposed by 3GPP is considered in this paper. Assuming that all MTDs are active between $t=0$ and $t=T_t$, the random access intensity is described by the distribution $f\left( t\right)$ and the total number of MTDs in the network, where the probability density function of distribution $f\left( t\right)$ is shown as follows \cite{37868}: 
\begin{align} f\left ( t \right )=\frac{t^{\alpha -1}\left ( T_t-t \right )^{\beta -1}}{T^{\alpha +\beta -1}\boldsymbol{B}\left ( \alpha ,\beta  \right ) }, t \in \left ( 0,T  \right ),   \label{e28} \end{align}
where $\boldsymbol{B}(\cdot )$ represents the Beta function and ($\alpha=3$, $\beta=4$) are considered in this paper. As the duration of each MAC frame is $\tau$, the number of newly activated MTDs at frame $i$ is given by 
\begin{align}v_{i}=M\int_{\left ( i-1 \right )\tau }^{i\tau} f\left ( t \right )dt, i=1,2, \cdots, T/\tau,    \label{e29}\end{align}
where $M$ is the total number of MTDs in the network. Note that the number of active MTDs in a frame (i.e., $\lambda$) is the summation of the number of newly active MTDs (i.e., $v_{i}$) and the number of backlogged active MTDs, where the backlogged active MTD refers to the MTD that suffered transmission failure in the previous frame and needs re-transmission in the current frame. 

\subsection{Throughput Performance } \label{section 5a}

\begin{figure}[t]
	\centering \includegraphics[height=6.5cm,width=8cm]{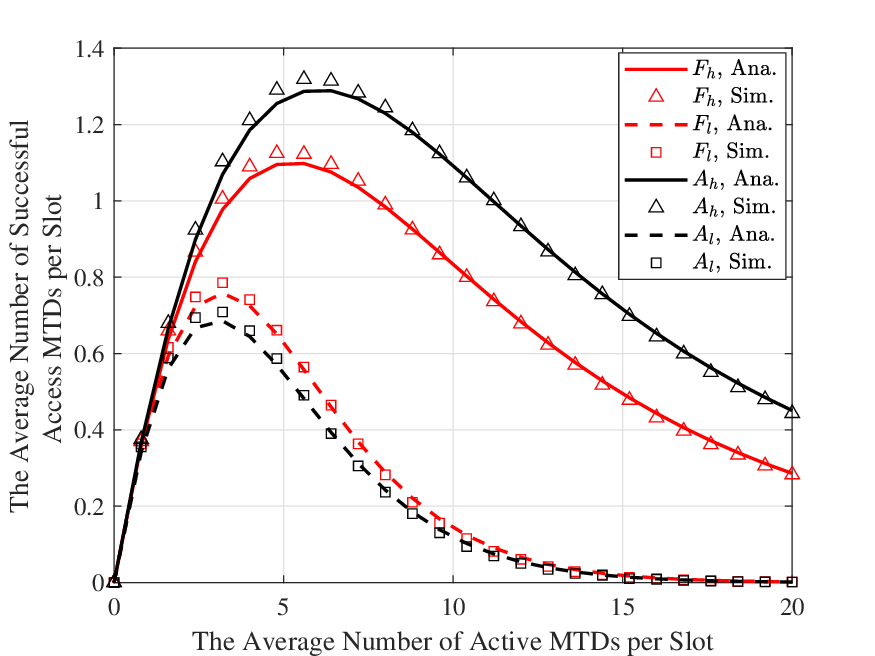}
	\caption{The average number of successfully accessed MTDs with different priorities per slot, when $Q=8$, $ \Gamma = 3$ dB, $\delta = 0.1\times\delta^*$, $\sigma_e^2=10^{-6}$, $\mu=0.5$, and $w=2$.}
	\label{f3}
\end{figure}
\begin{figure}[t]
	\centering \includegraphics[height=6.5cm,width=8cm]{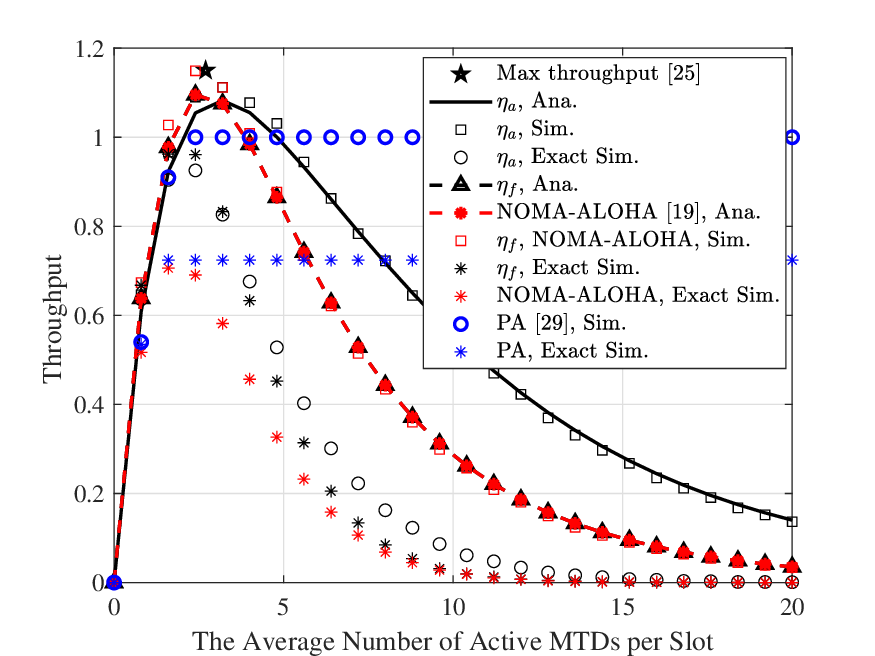}
	\caption{Throughput comparison of different schemes, when $Q=4$, $ \Gamma = 3$ dB, $\delta = 0.1\times\delta^*$, $\sigma_e^2=10^{-6}$, $\mu=0.5$, and $w=2$.}
	\label{f4}
\end{figure}
In Fig.~\ref{f3}, we present the theoretical upper bounds on the average number of successful accesses for high-priority and low-priority MTDs under the FPRA-NA and APRA-NA schemes, where $Q_1=Q_2$ is considered for the FPRA-NA scheme. It can be observed that the access performance of high-priority MTDs is significantly superior to that of low-priority MTDs. In the FPRA-NA scheme, although high-priority and low-priority MTDs are allocated the same number of high and low received power levels, a higher access success probability can still be guaranteed for high-priority MTDs. Therefore, the high-reliability transmission enabled by high received power levels plays a crucial role in improving the access performance of high-priority MTDs.

To verify the accuracy of the theoretical expressions, we also plot the Monte Carlo simulation results obtained under the same system design and assumptions adopted in the theoretical analysis. It can be seen that the Monte Carlo simulation results closely match the theoretical curves, thereby validating the correctness of the derived theoretical expressions.

Fig.~\ref{f4} presents the theoretical throughput upper bounds of the proposed PRA-NA schemes and the existing NOMA-ALOHA schemes, where $Q_1=Q_2$ is considered for the FPRA-NA scheme. It can be observed that the theoretical throughput of the proposed PRA-NA scheme is close to that of NOMA-ALOHA, but lower than the theoretical maximum throughput reported in \cite{10008538}. This is because the MTDs in the proposed scheme and NOMA-ALOHA uniformly select the received power levels, whereas the scheme in \cite{10008538} adopts an optimal power-level selection strategy to maximize the system throughput. Moreover, according to \eqref{e16}, when $Q_1=Q_2$ is adopted in the FPRA-NA scheme, its throughput upper bound coincides with that of the conventional NOMA-ALOHA scheme. To validate the theoretical expressions, we also plot the Monte Carlo simulation results obtained under the same system assumptions adopted in the corresponding analytical models. The consistency between the simulation results and the theoretical curves validates the correctness of the derived expressions.

\begin{figure}[t]
	\centering  \includegraphics[height=6.5cm,width=8cm]{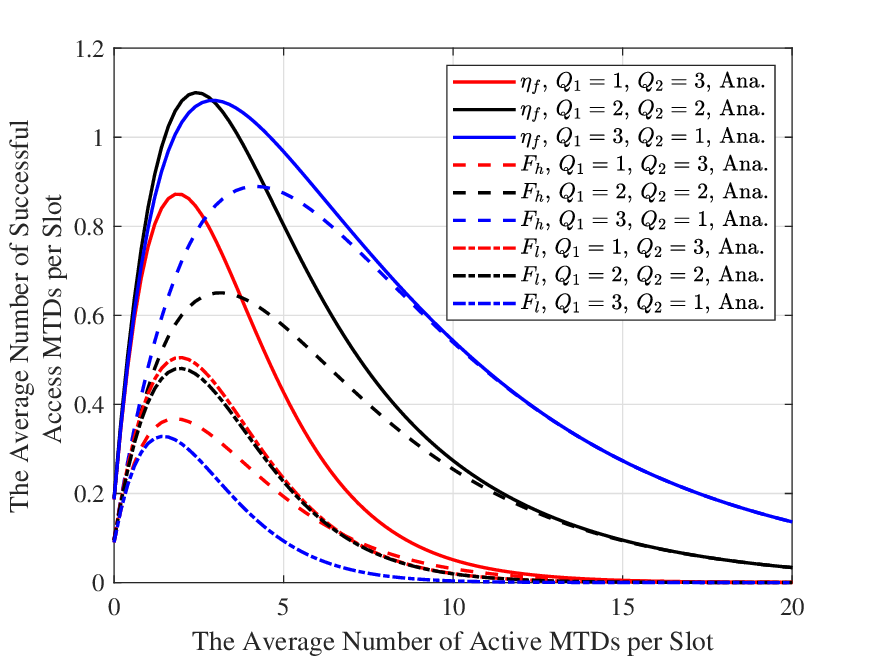}
	\caption{The performance of the FPRA-NA scheme with different $Q_1$ and $Q_2$, when $Q=4$ and $\mu=0.5$.}
	\label{f5}
\end{figure}

Furthermore, Fig.~\ref{f4} also shows the exact throughput performances of the proposed and existing schemes under imperfect SIC and imperfect CSI, whereby successful decoding requires the received SINR to exceed the target SINR threshold. These results are marked as ``Exact Sim.'' in the figure. In this case, power collisions with lower received power may affect the signal decoding at higher received power, which is ignored in the throughput upper-bound analysis. Therefore, the exact throughput is lower than the theoretical throughput upper bounds. Moreover, the proposed schemes achieve higher exact throughput than the conventional schemes, since the proposed received power level model explicitly accounts for the interference caused by imperfect SIC and imperfect CSI.

Fig.~\ref{f5} compares the performance of the FPRA-NA scheme for different values of $Q_1$ and $Q_2$. It can be observed that $F_h$ increases with $Q_1$, while $F_l$ increases with $Q_2$. Moreover, $\eta_f$ achieves better performance when the ratio $Q_1/Q_2$ is closer to the ratio of high-priority to low-priority MTDs.

\subsection{Average Successful Access Performance } \label{section 5b}

\begin{figure}[t]
	\centering    
	\subfigure[The proportion of high-priority MTDs $\mu$ increases, $Q=8$.] { \label{f6a}     \includegraphics[height=6.5cm,width=8cm]{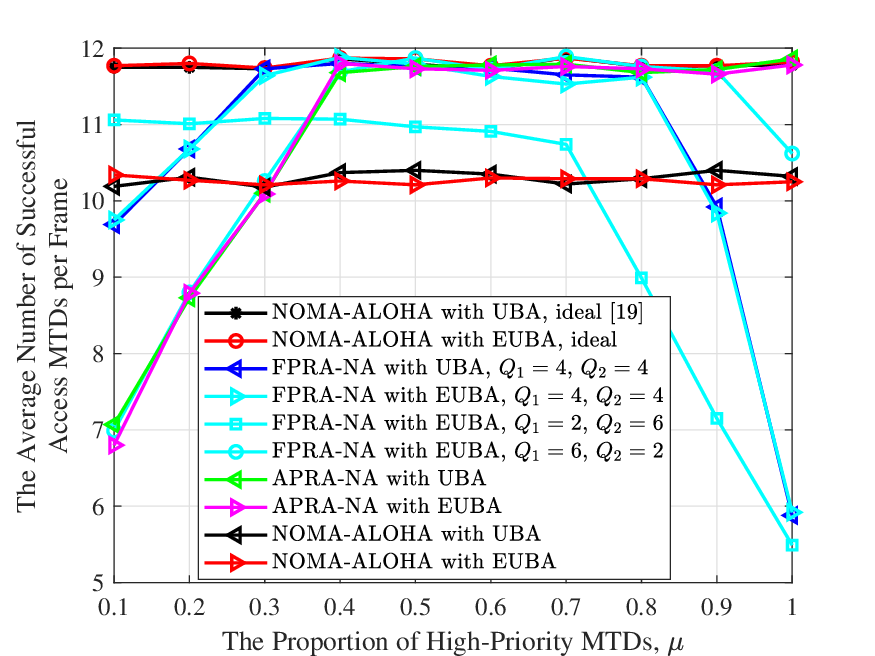}  } 
	\subfigure[The number of power level $Q$ increases, $\mu=0.5$.] { \label{f6b}  
		
	\includegraphics[height=6.5cm,width=8cm]{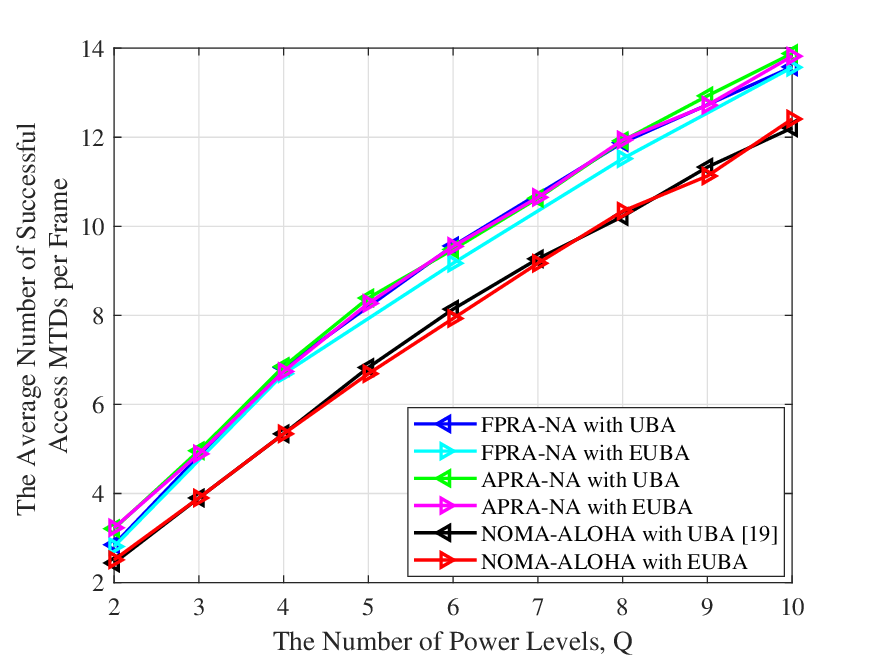}  }      
	\caption{ The average number of successful access MTDs per frame,  when $S=8$, $\Gamma=3$ dB, $w=2$, $\delta = 0.1\times\delta^*$, $\sigma_e^2=10^{-6}$, $\theta=0.5$, and $M=4000$. }     \label{f6}  
	
\end{figure}

\begin{figure}[t]
	\centering\includegraphics[height=6.5cm,width=8cm]{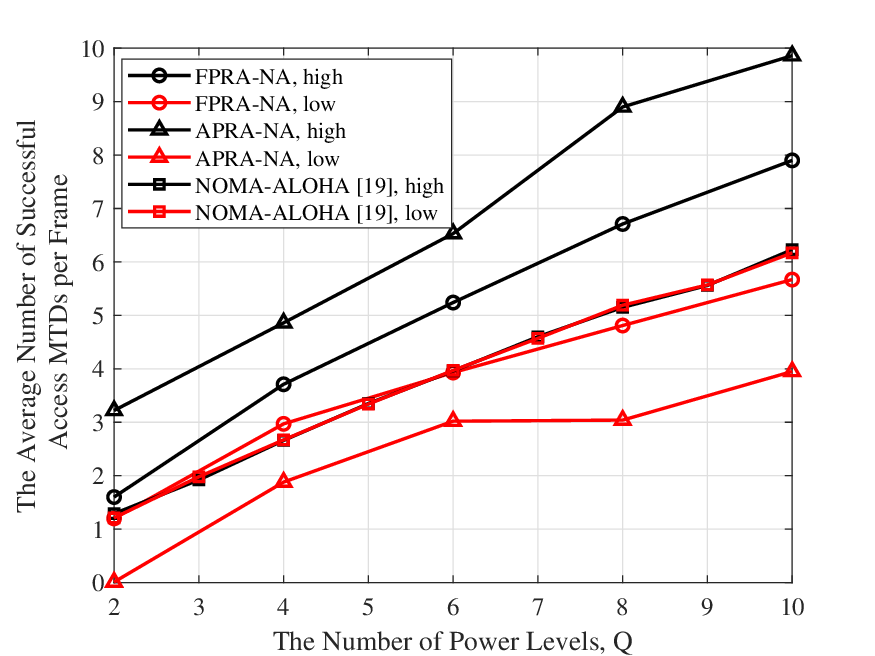}
	\caption{The average number of successful access of different priorities MTDs, when $S=8$, $\Gamma=3$ dB, $w=2$, $\delta = 0.1\times\delta^*$, $\sigma_e^2=10^{-6}$, $\mu=0.5$, $\theta=0.5$, and $M=4000$.}
	\label{f7}
\end{figure}

We carry out numerical studies to investigate the performance of the proposed PRA-NA schemes based on EUBA, where 3GPP Beta traffic is considered as the real-time traffic. The full details are given in Table II. The average number of successful access MTDs per frame under different schemes is displayed in Fig.~\ref{f6}. It can be observed that the performance of the schemes using EUBA is comparable to that of the conventional UBA, indicating that the proposed EUBA can effectively alleviate the user overload problem without performance loss. Fig.~\ref{f6a} illustrates the successful access performance versus the proportion of high-priority MTDs. It can be observed that imperfect CSI and imperfect SIC significantly degrade the performance of NOMA-ALOHA compared with the ideal case (i.e., perfect CSI and perfect SIC). In contrast, when the proportion of high-priority MTDs is large (e.g., $\mu > 0.4$), the proposed APRA-NA scheme achieves performance close to that of NOMA-ALOHA under ideal conditions. Similarly, for the FPRA-NA scheme, when the ratio $Q_1/Q_2$ becomes closer to the ratio of high-priority to low-priority MTDs, its performance approaches that of NOMA-ALOHA under ideal conditions.

Fig.~\ref{f6b} depicts the average number of successful access MTDs per frame as the number of power levels increases. It can be observed that the access performance improves with increasing $Q$. Moreover, the proposed PRA-NA schemes outperform the conventional NOMA-ALOHA scheme, since the latter is more sensitive to the interference caused by imperfect CSI and imperfect SIC.

Fig.~\ref{f7} shows the successful access performance of MTDs with different priorities as the number of power levels increases. It can be observed that, in the NOMA-ALOHA scheme, the successful access performance of high-priority MTDs coincides with that of low-priority MTDs, indicating that prioritized access cannot be supported. In contrast, under the proposed PRA-NA schemes, the access performance of high-priority MTDs consistently outperforms that of low-priority MTDs. Moreover, compared with the FPRA-NA scheme, the APRA-NA scheme further enhances the successful access performance of high-priority MTDs, because it allocates a larger number of high received power levels to high-priority MTDs, at the cost of reduced successful access performance for low-priority MTDs.

\begin{figure}[t!]
	
	\centering    
	\subfigure[The number of power level $Q$ increases, $\theta=0$.] { \label{f8a}     \includegraphics[height=6.5cm,width=8cm]{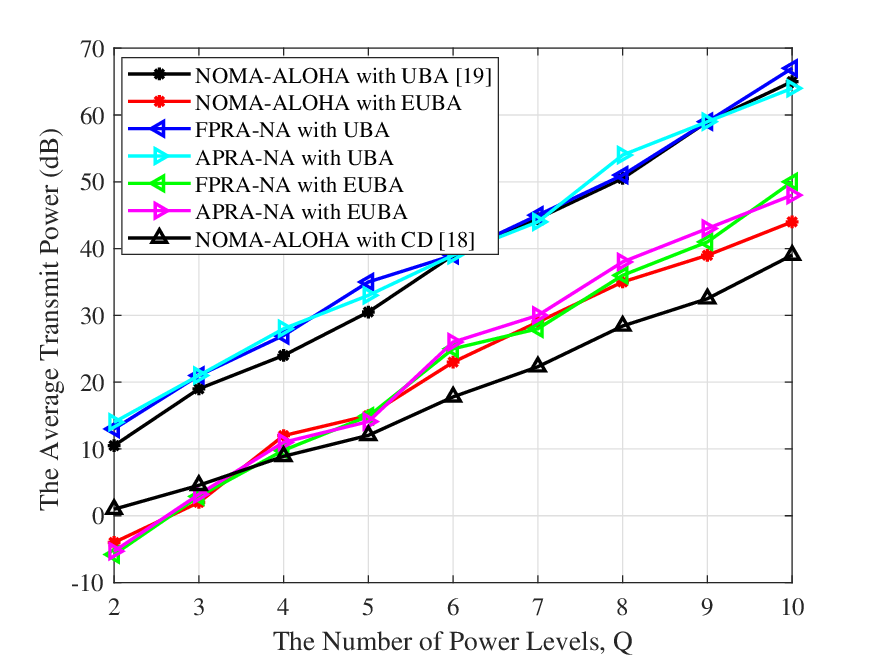}  }

	\subfigure[The trade-off factor $\theta$ increases, $Q=4$.] { \label{f8b}     \includegraphics[height=6.5cm,width=8cm]{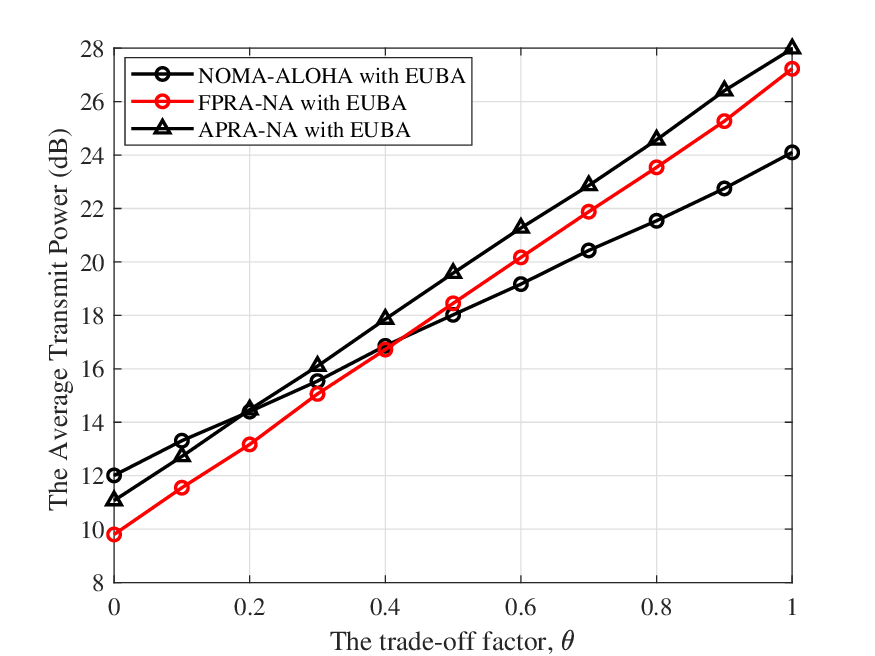}  }      
	\caption{ {The average transmit power per MTD that transmits data for different schemes, $S=8$, $\mu=0.5$, $w=2$, $\delta = 0.1\times\delta^*$, $\sigma_e^2=10^{-6}$, $\Gamma=6$ dB, and $M=4000$.} }     \label{f8}  
\end{figure}

\subsection{Average Transmit Power Performance } \label{section 5c}

Fig.~\ref{f8a} plots the average transmit power of the proposed PRA-NA schemes with EUBA, in comparison with the average transmit power achieved with conventional UBA and NOMA-ALOHA with CD scheme. It can be seen that the average transmit powers of the FPRA-NA and APRA-NA schemes using EUBA are lower than that of the schemes using UBA. This is due to the fact that the MTDs with poor channel quality are prohibited from transmission. However, when $Q$ is large (e.g., $Q>4$), the average transmit powers of the proposed FPRA-NA and APRA-NA schemes using EUBA are slight higher than the NOMA-ALOHA with CD scheme. The reason is that the high-power group is allocated more power levels, allowing even high-priority MTDs with poor channel quality to transmit data. Fig.~\ref{f8b} shows the impact of the trade‑off factor $\theta$ on the EUBA scheme. When $\theta$ = 1, the EUBA scheme reduces to the conventional UBA scheme, whereas when $\theta$ = 0, only the channel quality-based access control is applied to restrict MTD access. As $\theta$ increases, the weight of the conventional UBA in EUBA becomes larger, allowing more active MTDs to transmit data. Consequently, the effectiveness of the channel quality-based restriction is weakened, which leads to an increase in the average transmit power. Therefore, the capability of the EUBA scheme to reduce the average transmit power decreases with increasing $\theta$.

\begin{figure}[t]
	\centering\includegraphics[height=6.5cm,width=8cm]{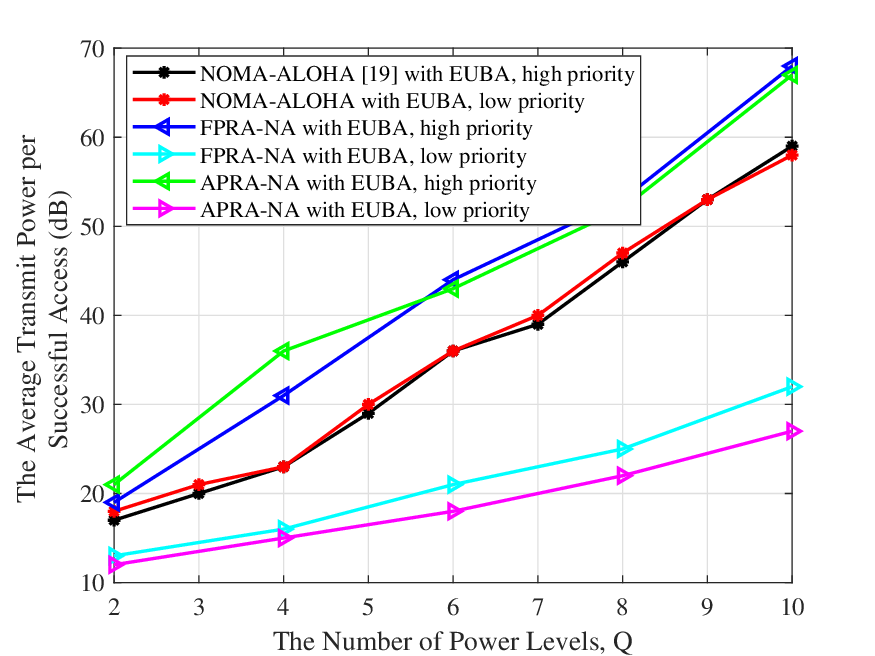}
	\caption{The average transmit power per successful transmission of different MTDs, $S=8$, $\mu=0.5$, $w=2$, $\delta = 0.1\times\delta^*$, $\sigma_e^2=10^{-6}$, $\Gamma=6$ dB, $\theta=0.5$, and $M=4000$.}
	\label{f9}
\end{figure}
Fig.~\ref{f9} illustrates the average transmit power per successful access for MTDs with different priorities. It can be observed that, under the proposed PRA-NA schemes (including FPRA-NA and APRA-NA), the average transmit power required by high-priority MTDs is higher than that in the NOMA-ALOHA scheme, while the average transmit power of low-priority MTDs is lower. This is because, in the PRA-NA schemes, high-priority MTDs are assigned higher received power levels to support prioritized access, whereas low-priority MTDs are allocated lower received power levels, resulting in reduced transmit power. These results indicate that the proposed PRA-NA schemes trade higher transmit power for higher throughput of high-priority MTDs.

\subsection{Average Access Delay Performance } \label{section 5d}
\begin{figure}[t!]
	\centering    
	\subfigure[The average access delay of MTDs] { \label{f10a}     \includegraphics[height=6.5cm,width=8cm]{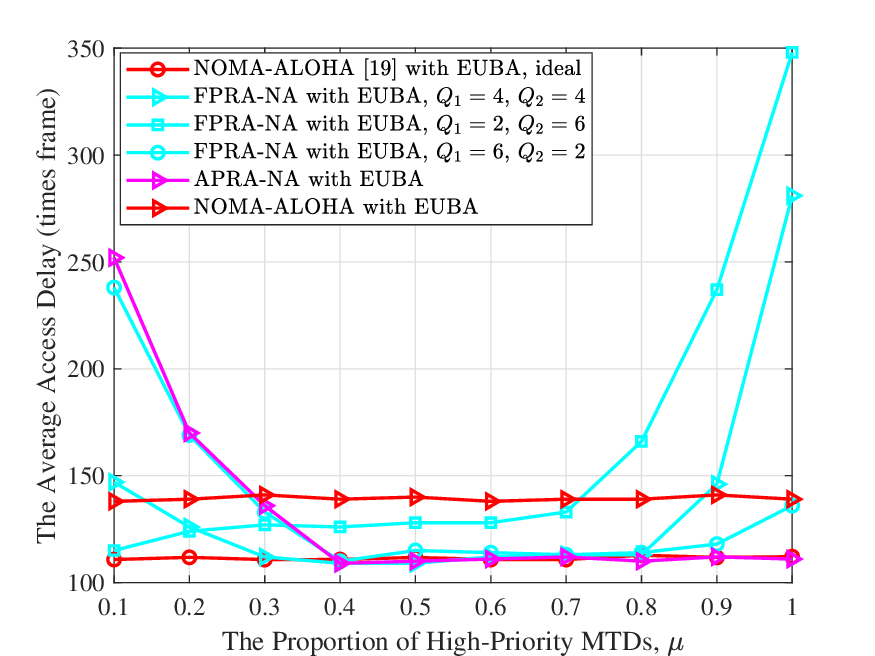}  }

	\subfigure[The average access delay of MTDs with different priorities.] { \label{f10b}     \includegraphics[height=6.5cm,width=8cm]{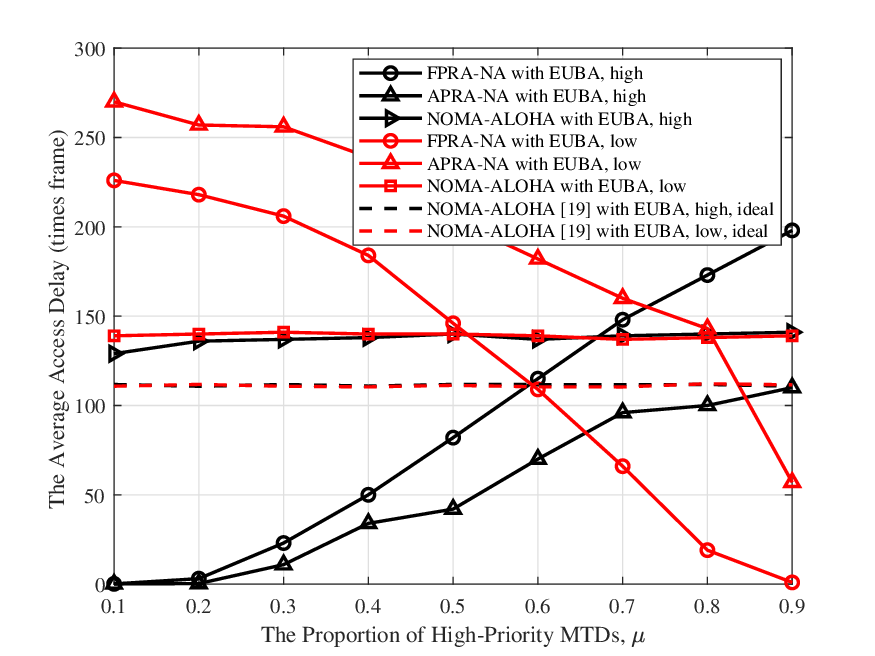}  }      
	\caption{ {The average access delay for different schemes, $S=8$, $\Gamma=3$ dB, $w=2$, $\delta = 0.1\times\delta^*$, $\sigma_e^2=10^{-6}$, $\Gamma=6$ dB, $\theta=0.5$, and $M=4000$.} }     \label{f10}  
\end{figure}

Fig.~\ref{f10} plots the average access delay for the proposed PRA-NA schemes and the conventional NOMA-ALOHA scheme. Fig.~\ref{f10a} presents the overall average access delay under different proportions of high-priority MTDs. Compared with the ideal NOMA-ALOHA case with perfect CSI and perfect SIC, the conventional NOMA-ALOHA scheme suffers from a noticeable delay increase when imperfect CSI and imperfect SIC are considered. This indicates that the interference caused by non-ideal channel estimation and SIC errors degrades the throughput performance, thereby significantly prolonging the average access delay.

For the proposed APRA-NA scheme, the average access delay remains close to the ideal NOMA-ALOHA benchmark when the proportion of high-priority MTDs is large, e.g., $\mu>0.4$. This is because APRA-NA can adaptively allocate more received power levels to high-priority MTDs, thereby improving the utilization of access resources under such traffic conditions. For the FPRA-NA scheme,  the average access delay performance is affected by the fixed allocation of $Q_1$ and $Q_2$. When $Q_1/Q_2$ is well matched to the traffic composition, i.e., the ratio of high-priority to low-priority MTDs, the access resources can be more efficiently utilized, leading to delay performance close to the ideal benchmark. Otherwise, mismatched power-level allocation may cause idle access resources and thus increase the average access delay.

\begin{figure*}[t!]
	\centering    
	\subfigure[ $\theta=1$.] { \label{f11a}     \includegraphics[height=6.5cm,width=8cm]{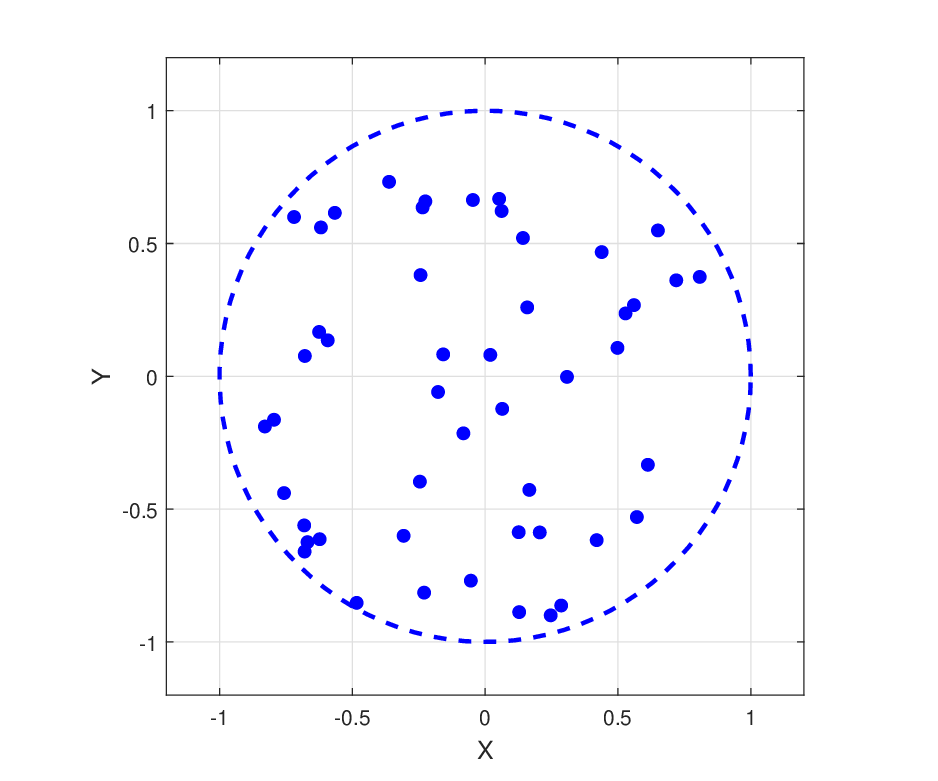}  } 
    \hspace{3em}
	\subfigure[$\theta=0$.] { \label{f11b}     \includegraphics[height=6.5cm,width=8cm]{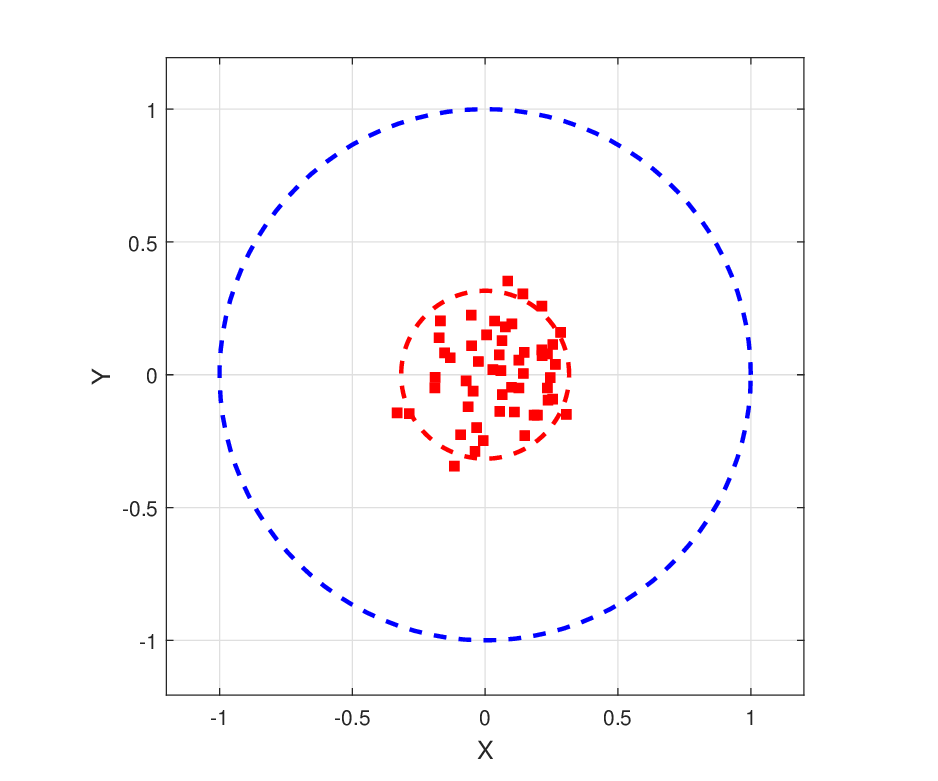}  }  

    \subfigure[$\theta=0.4$.] { \label{f11c}     \includegraphics[height=6.5cm,width=8cm]{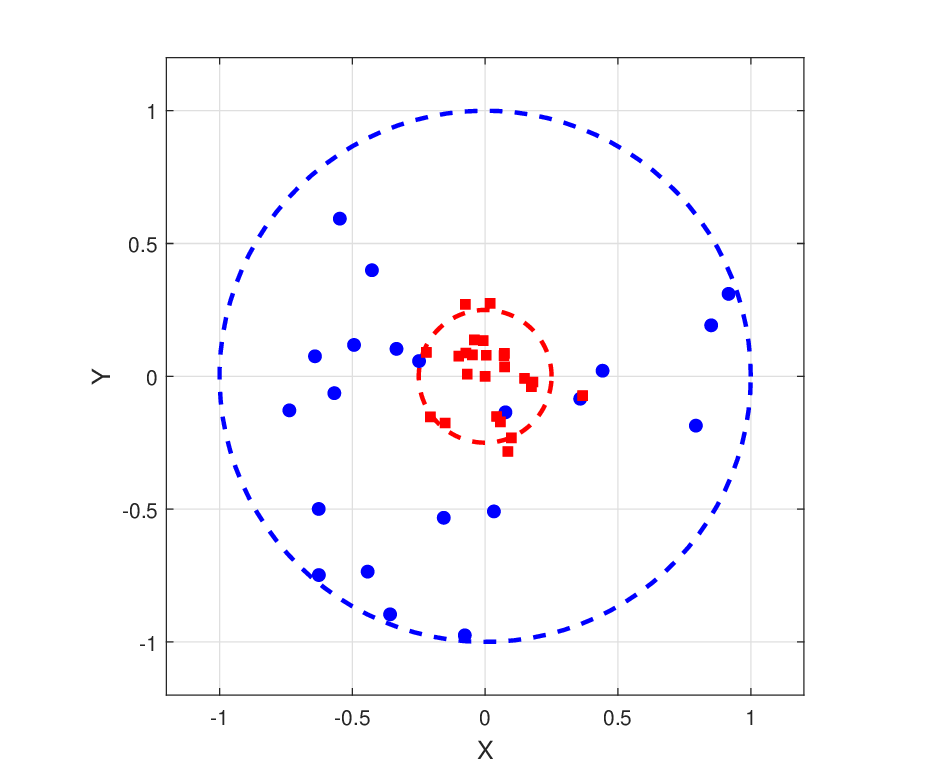}  }  
    \hspace{3em}
	\subfigure[$\theta=0.6$.] { \label{f11d}     \includegraphics[height=6.5cm,width=8cm]{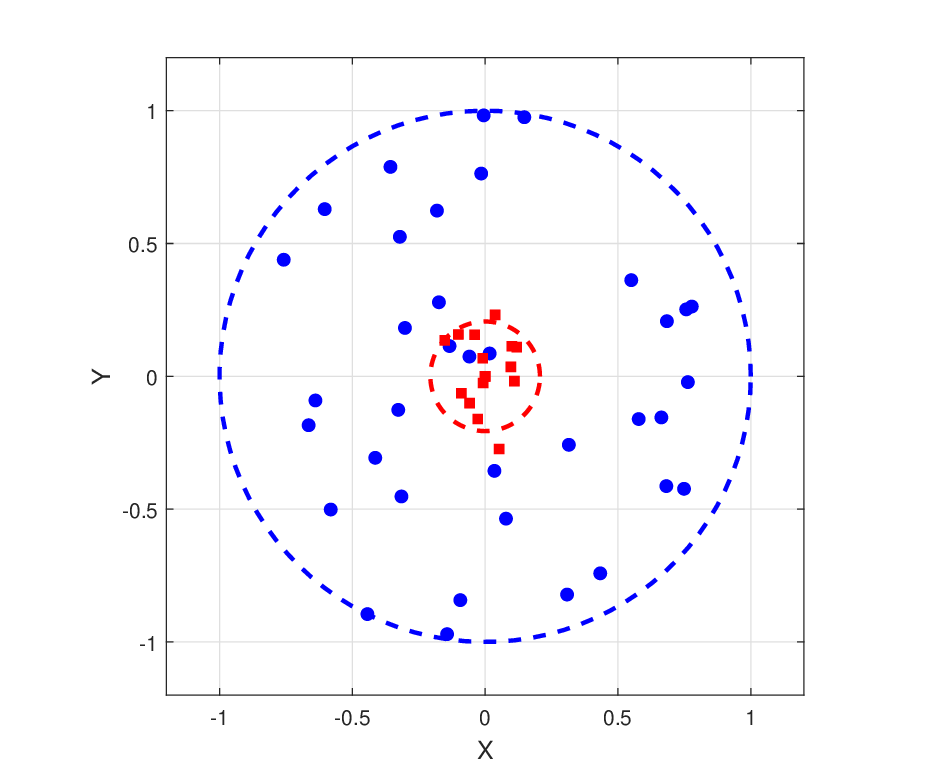}  }

	\caption{ {The location-dependent fairness for EUBA, when $\lambda=500$ and $P_b^*=0.1$.} }     \label{f11}  
\end{figure*}

Fig.~\ref{f10b} presents the average access delay performance of MTDs with different priorities. As the proportion of high-priority MTDs increases, the NOMA-ALOHA scheme maintains identical access delays for high-priority and low-priority MTDs under both ideal and non-ideal conditions, indicating that prioritized access cannot be supported. The proposed FPRA-NA scheme can reduce the access delay of high-priority MTDs when the number of high-priority MTDs is small (e.g., $\mu<0.6$); however, this advantage diminishes as $\mu$ increases. In contrast, the APRA‑NA scheme generally maintains a lower access delay for high-priority MTDs compared with the FPRA-NA and NOMA-ALOHA schemes.

Overall, combined with the results in Fig.~\ref{f7} and Fig.~\ref{f9}, a trade-off is revealed in the proposed PRA-NA schemes. The PRA-NA scheme achieves high throughput performance for high-priority MTDs by assigning them higher and more received power levels, at the cost of increased average transmit power. Meanwhile, the throughput performance of low-priority MTDs is relatively reduced. Since higher throughput generally leads to lower access delay, the PRA-NA scheme reduces the access delay of high-priority MTDs at the expense of degraded access delay performance for low-priority MTDs.

\subsection{Location-Dependent Fairness} 
Since the channel quality-based UBA scheme may reduce the effective coverage area of the eNB and thus compromise system fairness, Fig.~\ref{f11} illustrates the location-dependent fairness under different trade-off factors $\theta$. The blue points represent the active MTDs that transmit data after performing UBA, while the red markers indicate the active MTDs that transmit data after performing channel quality-based UBA. As shown in Fig.~\ref{f11a}, when $\theta=1$, the channel quality-based UBA is disabled and the EUBA reduces to the conventional UBA. In this case, the active MTDs permitted to transmit data are uniformly distributed within the cell, thus no location-dependent unfairness is observed. In contrast, when $\theta=0$ in Fig.~\ref{f11b}, the EUBA operates solely as the channel quality-based UBA. The active MTDs that transmit data are concentrated within the region defined by the channel quality threshold, leading to location-dependent unfairness, which may prevent MTDs far from the eNB from obtaining transmission opportunities. Therefore, a trade-off factor $\theta$ is required to mitigate such location‑dependent unfairness. As illustrated in Fig.~\ref{f11c} and Fig.~\ref{f11d}, when $\theta>0$, both the UBA and the channel quality-based UBA contribute to the EUBA scheme. Specifically, the UBA alleviates location-dependent unfairness by allowing MTDs far from the eNB to transmit data, while the channel quality-based UBA restricts transmissions from MTDs with poor channel quality, thereby reducing the average transmit power. In addition, as shown in Fig.~\ref{f11b}, Fig.~\ref{f11c}, and Fig.~\ref{f11d}, due to channel estimation errors, MTDs with actual channel quality below the threshold may be incorrectly admitted by the channel quality-based UBA procedure. Nevertheless, the resulting impact on both the system throughput and the average transmit power is negligible.

\section{Conclusion}\label{section 6}
In this paper, we studied priority-based random access and power control for NOMA-ALOHA in heterogeneous mMTC under non-ideal conditions. A received power level model considering imperfect CSI and imperfect SIC was developed, and PRA-NA schemes were designed to provide access priority for MTDs with different delay requirements under varying traffic conditions. Two PRA-NA strategies, namely FPRA-NA and APRA-NA, were proposed and their upper-bound throughputs were analyzed. Moreover, an EUBA was proposed and integrated with the PRA-NA schemes to alleviate the user overload problem and reduce the average transmit power.

Finally, several promising directions can be explored in future work. It is worth noting that, under certain extreme traffic scenarios, the proposed schemes may not achieve the maximum throughput due to the strict priority constraints. In such cases, a potential direction is to relax the priority requirements and jointly optimize the power-level allocation parameters (i.e., $Q_1$ and $Q_2$) with respect to throughput maximization. Moreover, the proposed scheme mainly focuses on access delay caused by MAC-layer retransmissions, while other delay components, such as physical-layer decoding and processing delays, were not considered. Nevertheless, the proposed scheme can naturally serve as a component of a holistic end-to-end delay optimization framework that incorporates cross-layer modeling.

\appendices
\section{}\label{appendix:A}
We assume that the eNB can successfully decode a data packet if its SINR is greater than or equal to $\Gamma$. Thus, the minimum received power for the $q$-th level can be written as
\begin{equation}
r_{q}=\Gamma(I_{q}+\Delta_{q}+r_q\sigma_e^2+\sigma^2), \tag{A1} \label{A1} 
\end{equation}
where $I_{q}=\sum_{i=q+1}^{Q}r_{i}$ represents the interference signals caused by the levels with lower received power,  $\Delta_{q}=\sum_{i=1}^{q-1}\delta r_{i}$ is the residual signals interference caused by the levels with higher received power due to the imperfect SIC. $r_q\sigma_e^2$ is the interference caused by the channel estimation error, and we assume that the channel estimation error is sufficiently small for the SIC technique to be applicable in the NOMA-ALOHA scheme. 

According to (\ref{A1}), we can obtain that
\begin{equation}
\begin{aligned}
r_q-r_{q+1} & =\Gamma\left( I_q-I_{q+1}+\Delta_q-\Delta_{q+1}+r_q\sigma_e^2-r_{q+1}\sigma_e^2\right)\\
&= \Gamma\left(r_{q+1}(1-\sigma_e^2)- r_q(\delta-\sigma_e^2)\right),
\end{aligned} \tag{A2} \label{A2}
\end{equation}
by further rearranging the terms, we obtain $\frac{r_q}{r_{q+1}}=\frac{1+(1-\sigma_e^2)\Gamma}{1+(\delta-\sigma_e^2)\Gamma}$, and $\{ r_Q,r_Q-1, \dots, r_1\}$ can be regarded as a geometric sequences with common ratio $\gamma=\frac{1+(1-\sigma_e^2)\Gamma}{1+(\delta-\sigma_e^2)\Gamma}$. Thus, the received power level model can be written as 
\begin{equation}
r_{q}=r_{Q}\left( \frac{1+(1-\sigma_e^2)\Gamma}{1+(\delta-\sigma_e^2)\Gamma}\right)^{Q-q}. \tag{A3} \label{A3} 
\end{equation}
According to (\ref{A1}), we have
\begin{equation}
\begin{aligned}
r_Q & =\Gamma\left( I_Q+\Delta_Q+r_Q\sigma_e^2+\sigma^2\right)\\
&= \Gamma\left(\delta \sum_{i=1}^{Q-1}r_i+ r_Q\sigma_e^2+\sigma^2\right),
\end{aligned} \tag{A4} \label{A4}
\end{equation}
according to the geometric sequences summation formula, we can obtain
\begin{equation}
\begin{aligned}
\sum_{i=1}^{Q-1}r_i & =\gamma\sum_{i=2}^{Q}r_i\\
&= \gamma\frac{r_Q\left(\gamma^{Q-1}-1 \right)}{\gamma-1}.
\end{aligned} \tag{A5} \label{A5}
\end{equation}
Thus, by substituting (\ref{A5}) into (\ref{A4}), $r_Q$ can be written as
\begin{equation}
\begin{aligned}
r_Q & =\Gamma\left(\delta\frac{r_Q\gamma\left( \gamma^{Q-1}-1\right)}{\gamma-1} +r_Q\sigma_e^2+\sigma^2 \right)\\
&= \frac{\Gamma \sigma^2}{1-\Gamma\sigma_e^2- \Gamma \delta \gamma\left(\gamma^{Q-1}-1\right)/\left(\gamma-1\right)},
\end{aligned} \tag{A6} \label{A6}
\end{equation}
where $1-\Gamma\sigma_e^2-\Gamma \delta \gamma\left(\gamma^{Q-1}-1\right)/\left(\gamma-1\right)>0$ due to $r_Q>0$. Consequently, the SIC efficiency is constrained by 
\begin{equation}
\delta  <\frac{(\gamma-1)(1-\Gamma\sigma_e^2)}{\Gamma \gamma (\gamma^{Q-1} - 1)}.  \tag{A7} \label{A7}
\end{equation}
To find the valid range for the SIC efficiency, define $f(\gamma)=\frac{(\gamma-1)(1-\Gamma\sigma_e^2)}{\Gamma \gamma (\gamma^{Q-1} - 1)}$, by taking the first derivative of $f(\gamma)$, we have
\begin{equation}
{f}'(\gamma) =(1-\Gamma\sigma_e^2)\frac{\left(Q+\gamma-\gamma Q\right)\gamma^{Q-1}-1}{\Gamma \gamma^2 (\gamma^{Q-1} - 1)^2},  \tag{A8} \label{A8}
\end{equation}
when $Q+\gamma-\gamma Q \le0$ and $1-\Gamma\sigma_e^2 >0$, by substituting $\gamma=\frac{1+(1-\sigma_e^2)\Gamma}{1+(\delta-\sigma_e^2)\Gamma}$, we obtain $\delta\le \frac{(Q-1)(1-\sigma_e^2)\Gamma-1}{Q\Gamma}$ and $\sigma_e^2<\frac{1}{\Gamma}$, in this case, $f(\gamma)$ is a monotonically decreasing function. Moreover, since $\delta \in (0,1)$, we know that $\gamma \in \left(1,1+\frac{\Gamma}{1-\Gamma\sigma_e^2}\right)$. When $\delta<f(1+\Gamma/(1-\Gamma\sigma_e^2))=\frac{1}{\left(1+\Gamma/\left(1-\Gamma\sigma_e^2\right)\right)^{Q}-\Gamma/\left(1-\Gamma\sigma_e^2\right)-1}$, the constraint in (\ref{A7}) can be satisfied due to that $f(\gamma)>f(1+\Gamma/(1-\Gamma\sigma_e^2))$. Therefore, the received power level model can be written as
\begin{equation}
r_{q}=\frac{\sigma^2\Gamma \gamma^{Q-q}(\gamma-1)}{(1-\Gamma\sigma_e^2)(\gamma-1)-\Gamma \delta \left(\gamma^{Q}-\gamma\right)}, \tag{A9} \label{A9} 
\end{equation}
where $\gamma=\frac{1+(1-\sigma_e^2)\Gamma}{1+(\delta-\sigma_e^2)\Gamma}$, and the SIC efficiency must satisfy that
$\delta<\min\left(\frac{(Q-1)(1-\sigma_e^2)\Gamma-1}{Q\Gamma}, \frac{1}{\left(1+\Gamma/\left(1-\Gamma\sigma_e^2\right)\right)^{Q}-\Gamma/\left(1-\Gamma\sigma_e^2\right)-1}\right). $ 

\vspace{1em}
\bibliographystyle{IEEEtran}
\bibliography{IEEEfull,TVT}
\end{document}